\documentclass[sigconf,screen,nonacm=true]{acmart}
\AtBeginDocument{\providecommand\BibTeX{{\normalfont B\kern-0.5em{\scshape i\kern-0.25em b}\kern-0.8em\TeX}}}

\usepackage{alltt}
\usepackage{amsmath}
\usepackage{paralist}           \usepackage{enumitem}
\usepackage{tikz, pgf}
\usepackage{tikzscale}
\usetikzlibrary{arrows.meta,decorations.pathmorphing,backgrounds,positioning,fit,petri}
\usepackage{graphicx}
\usepackage{standalone}   \usepackage{todonotes}
\usepackage{tikz}
\usetikzlibrary{arrows.meta,decorations.pathmorphing,backgrounds,positioning,fit,petri}

\usepackage{subfigure}  \usepackage{bm} \usepackage{listings}
\usepackage{xcolor}
\usepackage{textcomp} 

\usepackage{cleveref}
\crefformat{section}{\S#2#1#3}
\crefname{figure}{Figure}{Figures}
\crefname{table}{Table}{Tables}
\crefname{listing}{Listing}{Listings}
\crefname{assumption}{Assumption}{Assumptions}

\definecolor{keyword_color}{rgb}{0.8431372549019608, 0.0, 0.023529411764705882}
\definecolor{identifier_color}{rgb}{0.8980392156862745, 0.36470588235294116, 0.0}
\definecolor{string_color}{rgb}{0.1607843137254902, 0.1843137254901961, 0.40784313725490196}
\definecolor{number_color}{rgb}{0.027450980392156862, 0.396078431372549, 0.796078431372549}
\definecolor{annotation_color}{rgb}{0.43529411764705883, 0.23529411764705882, 0.7686274509803922}
\definecolor{grey}{rgb}{0.6, 0.6, 0.6}

\makeatletter
\lst@AddToHook{OnEmptyLine}{\vspace{-0.75\baselineskip}}
\newcommand*\idstyle{\expandafter\id@style\the\lst@token\relax
}
\def\id@style#1#2\relax{\ifcat#1\relax\else
                \ifnum`#1=\uccode`#1\color{number_color}
                \else
                        \color{identifier_color}
                \fi
        \fi
}
\makeatother

\lstdefinelanguage{graphql}{
  identifierstyle=\idstyle,
  delim=[s][\color{string_color}]{"}{"},morecomment=[f][\color{green!60!black}][0]{\#},
  literate=
     *{0}{{{\color{number_color}0}}}{1}
      {1}{{{\color{number_color}1}}}{1}
      {2}{{{\color{number_color}2}}}{1}
      {3}{{{\color{number_color}3}}}{1}
      {4}{{{\color{number_color}4}}}{1}
      {5}{{{\color{number_color}5}}}{1}
      {6}{{{\color{number_color}6}}}{1}
      {7}{{{\color{number_color}7}}}{1}
      {8}{{{\color{number_color}8}}}{1}
      {9}{{{\color{number_color}9}}}{1}
      {query\ }{{{\color{keyword_color}{query }}}}{1}
      {mutation\ }{{{\color{keyword_color}{mutation }}}}{1}
      {schema}{{{\color{keyword_color}{schema }}}}{1}
      {type\ }{{{\color{keyword_color}{type }}}}{1}
      {input\ }{{{\color{keyword_color}{input }}}}{1}
      {!}{{{\color{keyword_color}{!}}}}{1}
      {@deprecated}{{{\color{annotation_color}{@deprecated}}}}{1}
      {...\ on}{{{\color{keyword_color}{\texttt{... on }}}}}{1}
      ,
}

\lstnewenvironment{lstgql}{\lstset{
  language=graphql,
  basicstyle=\small\ttfamily,
  columns=fullflexible,
  xleftmargin=0em,
  belowskip=0em,
  aboveskip=0em,
  morecomment=[l][\color{olive}]{\#},
}}
{}
\def\gql{\lstinline[language=graphql, basicstyle=\small\ttfamily]}

\newcommand{\jstsize}[3]{\ensuremath{\texttt{tcx(} \mathit{#1}, #2, #3 \texttt{)}}}

\newcommand{\jsobject}[1]{\ensuremath{\texttt{\{} #1 \texttt{\}}}}
\newcommand{\jsfield}[2]{\ensuremath{\mathit{#1}: #2}}
\newcommand{\jslist}[2]{\ensuremath{\texttt{[}#1 \texttt{,} ... \texttt{,} #2 \texttt{]}}}

\newcommand{\jsaccess}[2]{\ensuremath{\mathit{#1}\texttt{.}\mathit{#2}}}
\newcommand{\jsget}[2]{\ensuremath{\mathit{#1}\texttt{[}\mathit{#2}\texttt{]}}}

\newcommand{\qestimate}{\ensuremath{\mathit{estimate}}}

\begin{document}

\title{Learning GraphQL Query Costs}
\subtitle{Extended Version}

\author{Georgios Mavroudeas}
\affiliation{\institution{Rensselaer Polytechnic Institute}
  \country{}
}

\author{Guillaume Baudart}
\affiliation{\institution{Inria Paris, {\'E}cole normale sup{\'e}rieure -- PSL university}
  \country{}
}

\author{Alan Cha}
\affiliation{\institution{IBM Research}
  \country{}
}

\author{Martin Hirzel}
\affiliation{\institution{IBM Research}
  \country{}
}

\author{Jim A. Laredo}
\affiliation{\institution{IBM Research}
  \country{}
}

\author{Malik Magdon-Ismail}
\affiliation{\institution{Rensselaer Polytechnic Institute}
  \country{}
}

\author{Louis Mandel}
\affiliation{\institution{IBM Research}
  \country{}
}

\author{Erik Wittern}
\affiliation{\institution{IBM Research}
  \country{}
}

\renewcommand{\shortauthors}{Mavroudeas, Baudart, Cha, Hirzel, Laredo, Magdon-Ismail, Mandel, and Wittern}

\newcommand{\remove}[1]{}

\begin{abstract}
  GraphQL is a query language for APIs and a runtime for executing those queries, fetching the requested data from existing microservices, REST APIs, databases, or other sources.
  Its expressiveness and its flexibility have made it an attractive candidate for API providers in many industries, especially through the web.
  A major drawback to blindly servicing a client's query in GraphQL is that the cost of a query can be unexpectedly large, creating computation and resource overload for the provider, and API rate-limit overages and infrastructure overload for the client.
  To mitigate these drawbacks, it is necessary to efficiently estimate
  the cost of a query \emph{before executing it}.
  Estimating query cost is challenging, because GraphQL queries have a nested structure, GraphQL APIs follow different design conventions, and the underlying data sources are hidden.
  Estimates based on worst-case static query analysis have had limited success because they tend to grossly overestimate cost.
  We propose a machine-learning approach to efficiently and accurately estimate the query cost.
  We also demonstrate the power of this approach by testing it on query-response data from publicly available commercial APIs.
  Our framework is efficient and predicts query costs with high accuracy, consistently outperforming the static analysis by a large margin.
\end{abstract}

\maketitle

\section{Introduction}\label{sec:intro}

GraphQL is an open-source technology for building APIs to support
client-server communication~\cite{fbcit, gql_foundation}.
GraphQL has two interconnected components: \begin{inparaenum}[(i)]
	\item a \emph{query language} that clients use to specify the data they want to retrieve or mutate, and
	\item a \emph{server-side runtime} to parse, validate, and execute these queries.
\end{inparaenum}

A central architectural design choice in GraphQL is to shift control over what data a request can receive or mutate from API providers to clients. 
In competing technologies, like the REpresentational State Transfer (REST) architecture, providers define accessible resources and their API endpoints.
In GraphQL, clients define queries that can retrieve or mutate multiple related resources in a single request 
(thus avoiding unwanted round-trips), and select only data they intend to use (thus avoiding over-fetching)~\cite{DBLP:journals/corr/abs-1906-07535, brito2020rest}.
As a result, GraphQL is very suitable for creating diverse client experiences and many organizations, such as Shopify, GitHub, Yelp, Starbucks, NBC, among others, 
have elected to use GraphQL to build mobile applications and engage with their ecosystem partners~\cite{graphql-users}.

Web API management is a challenging software engineering problem, for which
GraphQL provides advantages but also introduces new challenges.
A significant downside for providers when shifting control
to clients is the risk of overly complex queries, which are expensive
and lead to overloaded servers and/or databases.
Even small GraphQL queries can yield excessively large responses~\cite{hartig2018semantics,cha2020principled}. 
Empirical work shows that on many public GraphQL APIs,
a linear increase in query size can cause an exponential increase in result size
due to the nested nature of queries~\cite{wittern2019empirical}.

Unlike in REST APIs, where providers can avoid excessive API use by designing resources and endpoints carefully and limiting the number of allowed requests per time interval, in GraphQL, limiting the number of requests is not enough since
a single query can break the system.  
As such, some GraphQL server implementations track query \emph{costs}
dynamically during execution~\cite{graphQLsecurit}. 
Once a critical threshold is met, the server aborts execution and returns a partial result or an error. Unfortunately, this approach
can lock up resources while 
producing unusable results. 
Hartig et al.\ propose to analyze the cost of queries before executing them~\cite{hartig2018semantics}. 
Their analysis relies on probing the backend server for data-size information, for example, determining how many users are stored in the database if a query requests a list of users. 
However, this requires the server to offer probing facilities, which could themselves strain resources. 
In contrast, Cha et al.\ propose a static query cost analysis
that
does not depend on dynamic information from the server,
but in consequence only provides upper bounds
on cost~\cite{cha2020principled}.
This approach has been incorporated into IBM API Connect~\cite{APIConnect}, a commercially available API management product.

Unfortunately, these upper bounds are often loose and this
gap between estimated and actual cost makes the upper bound excessively
conservative as a
query filter, resulting in low amortized
efficiency.
More accurate cost estimates could allow providers to loosen their cost thresholds and help them better provision server resources.
In addition, clients can better understand the costs of their queries and how often they can execute them for given rate limits.

Therefore, we propose a machine-learning~(ML)
solution that 
predicts query costs based on experience generated over
multiple user-server communication sessions.
Our solution extracts features from query code by combining approaches
from natural-language processing, graph neural networks, as well as
symbolic features including ones from static compiler analysis
(such as the cost estimate in~\cite{cha2020principled}). It then
builds separate regressors for each set of features and combines
the component models into a stacking ensemble.

Compared to the static approaches, our solution can underestimate cost of a query but provides estimates that are closer to the actual value.

This paper makes the following contributions:
\begin{itemize}[leftmargin=*]
  \item A set of complementary feature extractors for GraphQL query
  code.
  \item A general ML workflow to estimate query cost
  that can be applied to any given GraphQL API.
  \item A search space of ML model architectures for GraphQL query cost
    prediction, comprising of choices for ensembling, preprocessing, and
    regression operators.
\item An empirical study of our approach on two commercial
    APIs, comparing it to previous
    work and evaluating the practical applicability.
\end{itemize}
Our approach can help API providers better
evaluate the risk of client queries, and it can help clients better
understand the cost of their queries to make the best use of their budget.

This paper is an extended version of~\cite{ase21}.

 \section{Background}\label{sec:background}

\begin{figure}[t]
  \centering
\begin{lstgql}
schema { query: Query }
type Query {
  viewer: User!
  licenses: [License]!
  repository(owner: String! name: String!): Repository }
type License { name: String! body: String! }
type Repository {
  issues(first: Int): IssueConnection!
  languages(first: Int): LanguageConnection }
type IssueConnection { nodes: [Issue] }
type LanguageConnection { nodes: [Language] }
type User { id: ID! name: String bio: String }
type Issue { id: ID! }
type Language { name: String! }
\end{lstgql}
  \caption{Simplified extract of the GitHub GraphQL schema.}
  \label{fig:schema}
\end{figure}

\begin{figure}[t]
  \centering
\begin{lstgql}
query {
  licenses { name }
  repository(owner: "graphql", name: "graphiql") {
    issues(first: 2) { nodes { id } }
    languages(first: 100) { nodes {name} } } }
\end{lstgql}
  \caption{Query for the GitHub GraphQL API.}
  \label{fig:query}
\end{figure}

\begin{figure}[t]
  \centering
\begin{lstgql}
{ "licenses": [
    {"name": "GNU Affero General Public License v3.0"},
    {"name": "Apache License 2.0"},
    {"name": "BSD 2-Clause \"Simplified\" License"},
    {"name": "BSD 3-Clause \"New\" or \"Revised\" License"},
    ... ],
  "repository": {
    "issues": {
      "nodes": [ {"id": "...NTQ="}, {"id": "...ODg="} ] },
    "languages": {
      "nodes": [ {"name": "HTML"}, {"name": "JavaScript"},
                  {"name": "Shell"}, {"name": "TypeScript"},
                  {"name": "CSS"} ] } } }
\end{lstgql}
  \caption{Response corresponding to the query of \Cref{fig:query}.}
  \label{fig:response}
\end{figure}

GraphQL queries are strongly typed by means of a \emph{schema}, and are executed via a set of data retrieval functions, called \emph{resolvers}.
The schema defines both the structure of queries and the types of the values returned. \Cref{fig:schema} is a simplified extract of GitHub's GraphQL API written with the Schema Definition Language~(SDL).
The entry point of the API is the \emph{field} \gql{query} which returns a value of type \gql{Query}.
A value of type \gql{Query} can contain the fields \gql{viewer}, \gql{licenses}, and \gql{repository} which return respectively a single \gql{User}~(the \gql{!} indicates that the value cannot be \gql{null}), a \emph{list} of \gql{License}s~(the square brackets \gql{[]} are list markers), and a \gql{Repository}.
The parentheses after the field \gql{repository} define additional arguments, indicating that the client must also provide the \gql{owner} and the \gql{name} of the requested repository.
GraphQL also includes a number of built-in scalar types such as \gql{String}, which is used in the field \gql{name} in \gql{Language}, and \gql{ID} used in the field \gql{id} in \gql{Issue}

Resolvers are functions that retrieve data for each field in an object type.
A resolver can obtain the data from any source, be it from a database, another API, or even from a file.

On the client side, the GraphQL \emph{queries} must follow the schema defined by the service provider.
Starting from the \gql{query} field, a query is a tree composed of nested fields such that the leaves of the tree must be fields of basic types~(\gql{Int}, \gql{String}, enumeration, etc.).
Fulfilling the query is a matter of calling the resolvers for each field in the query and composing the returned values into a response.
\Cref{fig:query} is an example of a valid query with respect to the schema of \Cref{fig:schema}.
This query asks for the list of open-source licenses available on GitHub and information about the \gql{"graphiql"} repository from the \gql{"graphql"} organization.
Notice that the query is composed of only desired fields and not all fields need to be requested (e.g. \gql{viewer} is not).
\Cref{fig:response} shows the response produced by the GitHub GraphQL API after executing the query of \Cref{fig:query}.
A JSON object is returned and it contains the same fields as the query.
For each field in the query with an object type~(e.g., \gql{repository}, which returns a value of type \gql{Repository}), the corresponding field in the response contains an object following the structure of the sub-query.
For each field in the query with a list return type (e.g., \gql{licenses}, which returns a list of \gql{License}s), the corresponding field in the response contains a list where each element is an object with the fields requested by the sub-query.

In order to reflect the cost of executing a query and the size of the response, Cha et al.\ define respectively the \emph{resolve complexity} and the \emph{type complexity}~\cite{cha2020principled}.
These complexities are the sums of the fields present in either the query or the response, weighted by a configuration associated to the type and resolver of each field.
To simplify the presentation, in this paper we only focus on type complexity.

Formally, we define the \emph{response type complexity}, $\jstsize{r}{t}{c}$, for a response~$r$ of type~$t$ with a configuration~$c$ as follows:
$$
\begin{array}{@{}l@{\,}c@{\,\,}l}
  \multicolumn{3}{@{}l}{
    \jstsize{\jsobject{\jsfield{field_1}{r_1}\texttt{,}...\texttt{,}\jsfield{field_n}{r_n}}}{t}{c} =
  }
  \\ && \jstsize{\jsfield{field_1}{r_1}}{\jsaccess{t}{field_1}}{c} +
  \\ && ... + \jstsize{\jsfield{field_n}{r_n}}{\jsaccess{t}{field_n}}{c}
  \\[.5em]

  \jstsize{\jsfield{field}{v}}{t}{c} & = &
  \jsaccess{\jsget{c}{t}}{\texttt{typeWeight}}
  \\ &&\text{when } v \text{ is a scalar}
  \\[.5em]

  \jstsize{\jsfield{field}{r}}{t}{c} & = & \jsaccess{\jsget{c}{t}}{\texttt{typeWeight}} + \jstsize{r}{t}{c}
  \\ &&\text{when } r \text{ is an object}
  \\[.5em]

  \jstsize{\jsfield{field}{\jslist{r_1}{r_n}}}{t}{c} & = &
  w + \jstsize{r_1}{t}{c}\ +
  \\ && ... + w + \jstsize{r_n}{t}{c}
  \\ &&\text{where } w = \jsaccess{\jsget{c}{t}}{\texttt{typeWeight}}
\end{array}
$$

The case $\jstsize{\jsobject{\jsfield{field_1}{r_1}\texttt{,}...\texttt{,}\jsfield{field_n}{r_n}}}{t}{c}$ is the entry point of the function: if the response is a JSON object composed of the fields $\mathit{field}_1$, ..., $\mathit{field}_n$ then the complexity is the sum of the complexity of each field.
The other three cases correspond to the definition of the complexity of a field depending on the shape of the value associated to it.
$\jstsize{\jsfield{field}{v}}{t}{c}$ corresponds to the case where the value $v$ is a scalar (number, string, or boolean). The complexity is the weight of the type $t$~(which is the type of the field $\mathit{field}$) in the configuration $c$.
$\jstsize{\jsfield{field}{r}}{t}{c}$ corresponds to the case where $r$ is a JSON object. The complexity is the weight associated to the type of the field in the configuration plus the complexity of the object $r$.
Finally, $\jstsize{\jsfield{field}{\jslist{r_1}{r_n}}}{t}{c}$ corresponds to the case where the value of the field is an array of objects. The complexity is $n$ times the weight of the type $t$ plus the sum of the complexities of every elements of the array.

If we use a configuration where the weight of a scalar type is~0, and the weight of all other types is 1, then the type complexity of the response in \Cref{fig:response}, with 13 \gql{"licenses"} and 5 \gql{"languages"}, is~23~(= 13~\gql{License}s + 1~\gql{Repository} + 1~\gql{IssueConnection} + 2~\gql{Issue}s + 1~\gql{LanguageConnection} + 5~\gql{Language}s).

Without list types, the cost of a query will always be, at worst, linear with respect to the size of the query because each field in the query should have a corresponding field in the response.
Thus, the issues of estimating query costs come from the lists which can be of arbitrary length.
Moreover, nested lists can yield exponentially large responses~\citep{hartig2018semantics}.
In our example, the length of the lists \gql{issues} and \gql{languages} are bounded by the argument \gql{first}, as dictated by the connection model~\citep{connection-model}.
Cha et al.\ \citep{cha2020principled} use this information to statically compute an upper bound on the response size from the query.
While the upper bound they compute is as tight as possible, the estimate (known as \emph{query type complexity}) can grossly differ from the actual cost~(known as \emph{response type complexity}).
For example, the static analysis assumes that the query of \Cref{fig:query} returns at worst a list of 100 programming languages, but the GraphiQL repository uses only 5~programming languages.

\section{Methodology}
\label{sec:problem}

The goal of this work is to automatically learn more accurate query cost estimates from data.
First, we propose a set of specialized features that can be applied to any GraphQL API. 
These features turn GraphQL queries into suitable input for classic machine learning techniques.
Second, we propose a hierarchical model to learn a cost estimate given a GraphQL query.
Separate regressors for each features are combined into a stacking ensemble to obtain the final estimate.

\subsection{Setup}

Let $\mathcal{Q}$ and $\mathcal{R}$ be the space of the possible queries and responses, $\mathit{SDL}$ the set of possible schemas\footnote{SDL: Schema Definition Language}, and $\mathcal{C}$ the set of possible configurations.
Given a valid GraphQL schema $s \in \mathit{SDL}$ and a series of $n$ query-response pairs ${(q_i, r_i) \in \mathcal{Q} \times \mathcal{R}}$ with ${i \in \{1,\ldots, n\}}$, we would like to learn a function ${\qestimate: \mathit{SDL} \times \mathcal{C} \times \mathcal{Q} \rightarrow \mathbb{R}}$ which returns an estimation of the type complexity of a query $q$ given a schema~$s$, and a configuration~$c$:
$$
\qestimate(s, c, q) \approx \jstsize{r}{\mathtt{Query}}{c}
$$

To simplify the problem, we decompose $\qestimate(.)$ into two functions ${h: \mathit{SDL} \times \mathcal{C} \times \mathcal{Q}\rightarrow \mathbb{R}^k}$ and ${m: \mathbb{R}^k \rightarrow \mathbb{R}}$:
$$
\qestimate(s, c, q) = m(h(s, c, q))
$$
The function $h$ defines an \emph{embedding} of GraphQL queries. Given a schema a configuration and query which adhere to the schema, it returns a numerical vector representing the query.
Then $m$ takes a query embedding and returns an estimate of the complexity.

 \subsection{Feature Extraction}
\label{sec:features}

We design three distinct feature extraction methods.

\paragraph*{Field Features}

A GraphQL schema defines a finite number of types and a finite number of fields.
We can thus represent all possible fields by a vector of fixed size where each index represents a particular field.
We create a feature vector for each GraphQL query, enumerating the total number of times a specific field appears inside a query.
This corresponds to a Bag-of-Words representation of the query~\cite{harris54}.
We call this feature extraction function~$h_f$.

\paragraph*{Graph Embeddings}

The field features only capture information about the cardinality of fields.
To capture information about the syntactic structure of the query, we use a second set of features based on graph embeddings.
The idea is that a graph neural network can map the abstract syntax tree of a GraphQL query into a low-dimensional embedding space, from which we can then extract the numerical features.
To do that, we employed the graph2vec~\citep{narayanan2017graph2vec} technique, one of the most popular approaches in this area.
The idea of graph2vec is to define a function that takes as input a graph and returns a vector of real numbers such that similar graphs are mapped to similar vectors.
To define this function, a neural network is trained.
The principle of the training is the same as the one of doc2vec where documents are graphs and words are sub-graphs.
The objective of the training is to put next to each other in the embedding space the sub-graphs that appear in the same context and propagate this information at the graph level.
The training does not need labeled data, it just needs a set of graphs.
Once trained, to compute the embedding, a graph is decomposed into a set of sub-graphs and fed to the neural network. The value of a low dimension hidden layer of the network is used as the vector characterizing the graph.
We call this feature extraction function~$h_g$.

\paragraph*{Summary Features}
The last set of features is a six-dimensional encoding of the queries using symbolic code analysis techniques.
They include
\begin{inparaenum}[(i)]
\item the \emph{static analysis upper bound} of Cha et al.\ \citep{cha2020principled}.
They also include features that summarize the structure of the abstract syntax tree of the queries.
These are
\item \emph{query size}, the number of nodes in the abstract syntax tree,
\item \emph{width}, the maximum number of children a tree node has, and
\item \emph{nesting}, the maximum depth of the tree.
Finally, we extract two features related to lists:
\item \emph{lists}, the number of fields in a query requesting a list, and
\item the \emph{sum of list limits} (e.g, \gql{first})\end{inparaenum}.
The features vector of \Cref{fig:query} is $[ 118, 17, 2, 3, 3, 115]$
(the list \gql{licenses} has default length~13).
We call the function that extracts all of these summary features~$h_s$.

 \subsection{Learning}
\label{sec:learning}

There are many well-known \emph{operators} that implement regression algorithms (e.g., linear regression, gradient boosting regressors) and also feature preprocessing (e.g., polynomial features transformer).
A library like scikit-learn~\citep{scikit-learn} implements many of these operators, but picking the right operators and configuring their hyperparameters is a tedious task and depends on the dataset.
We thus use Lale \citep{baudart_et_al_2020-automl_kdd}, a state-of-the-art automated machine learning (auto-ML) tool, to automatically select the best operators and tune the hyperparameters given a query/response dataset.

\paragraph*{Definition of three models}

\begin{figure}
  \centering
  \includegraphics[width = 0.6\linewidth]{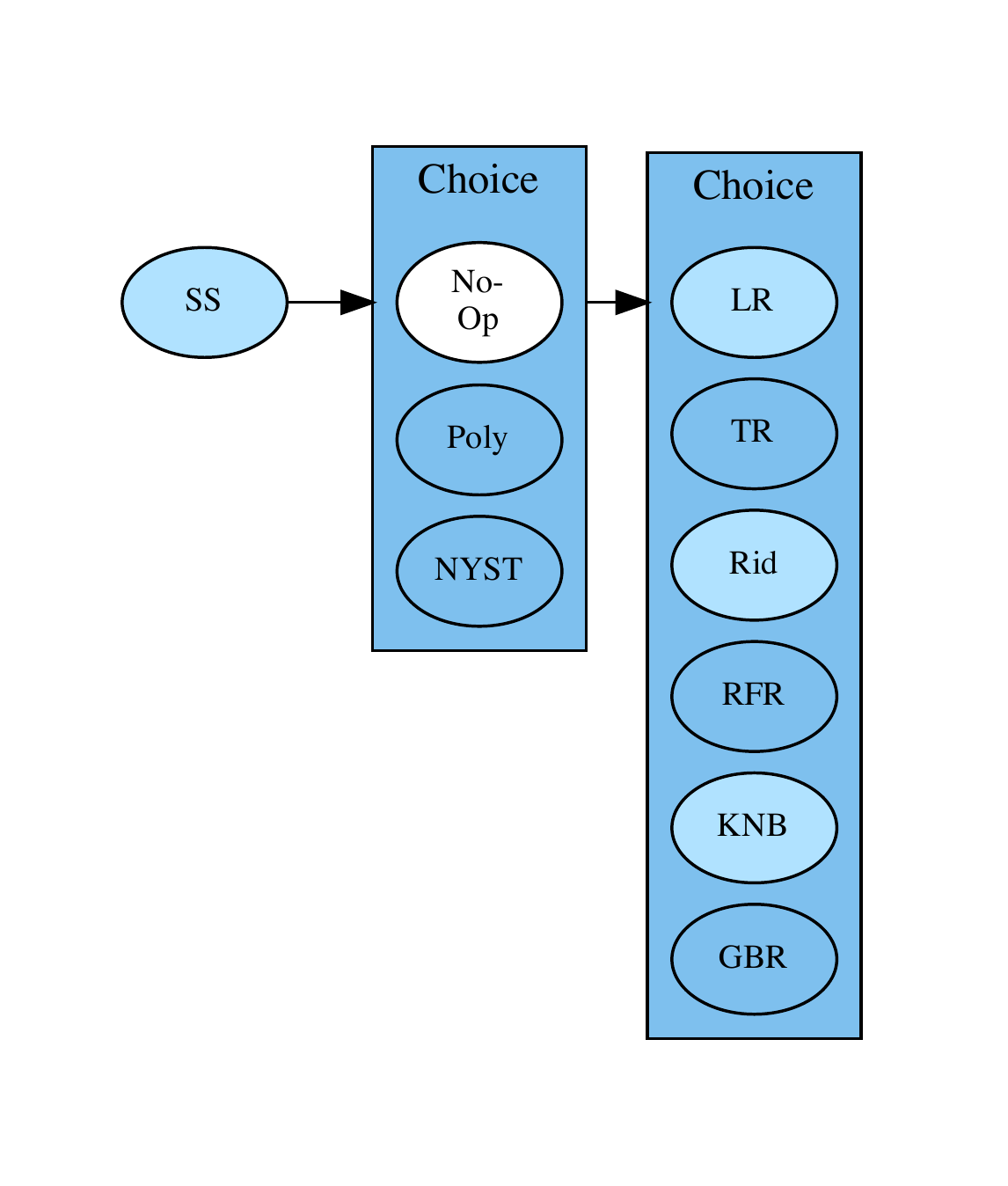}
  \caption{Lale pipeline: round nodes represent ML operators and rectangular nodes represent a choice between different operators.}
  \label{fig:pipeline}
\end{figure}

In Section~\ref{sec:features}, we defined three set of features:
\textit{field features},
\textit{graph embeddings features}, and
\textit{summary features},
For each set of features, we train a model independently, but all three of these models have the same architecture presented in \Cref{fig:pipeline}.
We define a pipeline where
\begin{inparaenum}[(i)]
\item the first step uses a standard scaler (\textsf{SS}) to give all features a mean of~$0$ and a standard deviation of~$1$;
\item the second step does other feature transformation; and
\item the last step does the prediction.
\end{inparaenum}

For the feature transformation and prediction steps, we configure the auto-ML tool such that it chooses the best solution among multiple algorithms.
There are three possible feature transformations:
\textsf{No-Op} leave the features unchanged,
\textsf{Poly} is a polynomial expansion of the features,
and \textsf{NYST} is Nystroem transformer.
For the last part, we selected six different predictors:
linear regression~(\textsf{LR}),
decision tree~(\textsf{TR}),
Ridge regression~(\textsf{Rid}),
random forest~(\textsf{RFR}),
$k$-nearest neighbors where the number of neighbors is fixed to~3~(\textsf{KNB}), and
gradient boosting regressors~(\textsf{GBR}).
\begin{table}
  \centering
  \begin{tabular}{l@{ }lrrrr}
    & & \multicolumn{2}{c}{Categorical} & \multicolumn{2}{c}{Continuous} \\
    \cmidrule(l){3-4}\cmidrule(l){5-6}
    \multicolumn{2}{l}{Operator} & free & total & free & total \\
    \midrule
    \textsf{SS}    & (standard scaler)      & 0 & 2 & 0 & 0  \\
    \textsf{No-Op} & (identity transform)   & 0 & 0 & 0 & 0  \\
    \textsf{Poly}  & (polynomial features)  & 3 & 3 & 3 & 3  \\
    \textsf{NYST}  & (Nystroem)             & 3 & 3 & 2 & 2  \\
    \textsf{LR}    & (linear regression)    & 0 & 2 & 0 & 0  \\
    \textsf{TR}    & (decision tree)        & 4 & 4 & 2 & 2  \\
    \textsf{Rid}   & (Ridge regression)     & 3 & 5 & 2 & 2  \\
    \textsf{RFR}   & (random forest)        & 5 & 5 & 2 & 2  \\
    \textsf{KNB}   & ($k$ nearest neigbors) & 4 & 5 & 0 & 0  \\
    \textsf{GBR}   & (gradient boosting)    & 8 & 8 & 2 & 2  \\
    \bottomrule
  \end{tabular}
  \vspace{0.5em}
  \caption{Number of hyperparameters to optimize and total for each operator in the pipeline of \Cref{fig:pipeline}.}
  \label{fig:hyperparameters}
\end{table}

\paragraph*{Model selection}
This pipeline defines a space of 18 possible combinations for each of the three models.
Furthermore, each of the operators of the pipeline also has a set of hyperparameters to configure.
We have fixed some of the hyperparameters, such as $k=3$ for the $k$-nearest neighbors, but we left 43 hyperparameters free.
The auto-ML tool then chooses the best solution among the possible combinations of algorithms and hyperparameters configurations.
To select the best model, we use $n$-fold cross validation and the Bayesian optimizer from Hyperopt~\citep{hyperopt-sklearn}.

\paragraph*{Combination of models}

We train the three models independently and define a new hierarchical model using the outputs of the three models as input for a final model.
This final model provides the final estimation in the prediction phase.
In general, using a \emph{stacked} ensemble in an ML framework~\citep{wolpert1992stacked}, learning each predictor separately and using the predictions as features for the final predictor, can improve accuracy.
After experimentation, we found that in our case, this method performs better than concatenating the individual features together into a wide vector and using a single regressor.

\begin{figure}[t]
	\begin{tikzpicture}[
					node distance=1.1cm,
					cir/.style={circle, inner sep=0, minimum size=8mm,fill=white!80!gray, thin, draw = white!50!gray},
					rec/.style={rectangle, inner sep=0,minimum height = 6mm, minimum width=16mm,fill=white!80!gray, thin, draw = white!50!gray},
					pre/.style = {<-,shorten <=1pt, >={Stealth[round]}, semithick},
	   			    post/.style={->,shorten >=1pt, >={Stealth[round]}, semithick}]

\node[cir] (q){$q$};
	 \node[cir, below= of q, xshift=-1.5cm] (xf){$x_f$};
	 \node[cir,below= of q] (xi){$x_g$};
	 \node[cir, below= of q, xshift=1.5cm] (xem){$x_s$};
	 \node[cir, below=of xi, xshift = -1.5cm] (c1){$\hat{c}_f$};
	 \node[cir, below= of xi] (c2){$\hat{c}_g$};
	 \node[cir, below= of xi, xshift = 1.5cm] (c3){$\hat{c}_s$};
	 \node[cir, below of= c2, yshift=-0.5cm, scale = 0.75] (mf){$[\hat{c}_f\hat{c}_g\hat{c}_s]$};
     \node[cir, below= of mf] (cf){$\hat{c}_\mathit{final}$};

\draw (xf)  edge[pre, bend left=45] node[auto]{$h_f(\cdot)$} (q)
     			 edge[post]node[swap, xshift=-6mm]{$m_f(\cdot)$}(c1);

     \draw (xi)  edge[pre] node[auto]{$h_g(\cdot)$} (q)
                 edge[post]node[swap, xshift=-6mm]{$m_g(\cdot)$}(c2);

     \draw (xem) edge[pre, bend right=45] node[auto, swap]{$h_s(\cdot)$} (q)
                 edge[post]node[swap, xshift=-6mm]{$m_s(\cdot)$}(c3);

     \draw (mf)  edge[pre]  (c1)
    		     edge[pre](c2)
                 edge[pre](c3);

     \draw (mf) edge[post] node[auto, swap]{$m_{\mathit{final}}(\cdot)$}
     (cf);

	\end{tikzpicture}
     \caption{Cost estimation process for an input query $q_i$. First, we extract the numerical features $x_f, x_g, x_s$, from the corresponding functions
    $h_f, h_g, h_s$. The features are inserted into the independent cost estimation models $m_f, m_g, m_s$, which in turn produce the first cost estimates for each of the features. These costs are concatenated into a new feature vector $[\hat{c}_f, \hat{c}_g, \hat{c}_s]$, the input for the final model $m_{\mathit{final}}$ which generates the final cost estimate $\hat{c}_\mathit{final}$.}
	\label{fig:process}
\end{figure}
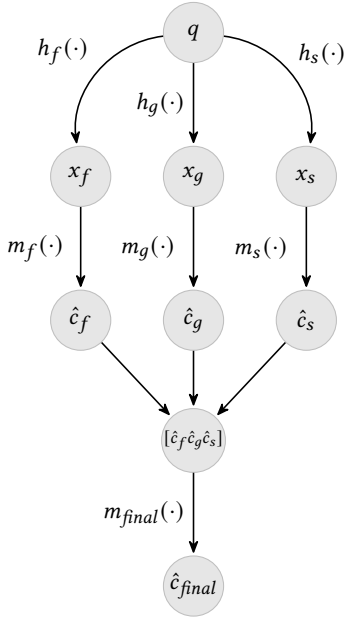

\Cref{fig:process} describes our final architecture.
More formally, given a training set of query-response pairs $X = \{(q_1,r_1),\ldots, (q_n, r_n)\}$, 
we train three distinct models $m_f, m_g, m_s$, each taking as inputs the corresponding features sets we constructed from the queries earlier.
Each model returns a cost estimation $\hat{C}_f$, $\hat{C}_g$, or $\hat{C}_s$, which will be an $m$-dimensional vector~(the cost estimation for each of the queries in the training set).
In turn we use these three features to train a new model $m_\mathit{final}$, which returns the final cost prediction $\hat{C}_\mathit{final}$.
After the learning phase, the cost prediction on a new query $q$ can be done in a similar way.
First, we extract the features $x_f, x_g, x_s$, from the query, then we get the cost estimates $\hat{c}_f, \hat{c}_g, \hat{c}_s$ from $m_f, m_g, m_s$ respectively, and finally we insert these estimates into our last model $m_{\mathit{final}}$ to get the final cost estimate $\hat{c}_\mathit{final}$.

 \section{Results}
\label{sec:results}

In our evaluation, we first study how our new ML query cost estimation compares to both the static analysis in~\cite{cha2020principled} and the actual cost of the response.
We then evaluate the importance of the different features that we have extracted to obtain a good estimate of the cost.
Finally, we study the practical usage of our approach.
For that, we look into how an API management layer could filter queries using the ML estimates and simulate how our estimator would react to malicious queries.

This can be summarized with the following research questions:
\begin{description}
\item[RQ1:] Does our approach return accurate estimates?
\item[RQ2:] Are all the features useful for the estimation?
\item[RQ3:] What are the practical benefits of the new estimation?
\item[RQ4:] Are the ML cost estimates robust to malicious queries?
\end{description}

\subsection{Experimental Setup}
\label{sec:experimental-setup}

\paragraph{Data}

Following the methodology in~\cite{cha2020principled}, we collected over \mbox{100,000} responses from the GitHub GraphQL API and \mbox{30,000} from the Yelp GraphQL API.
The queries are synthetically generated but the responses come from real industry APIs.
The dataset is available at \url{https://github.com/Alan-Cha/graphql-complexity-paper-artifact}.
Table~\ref{table:stats} presents the dataset characteristics, where
\emph{query size}, \emph{width}, \emph{nesting}, and \emph{lists} are
the corresponding summary features from Section~\ref{sec:features},
and \emph{response} is the actual cost of the query result.

\begin{table}
\caption{\label{table:stats}Data statistics for the GitHub and Yelp datasets.}
\centerline{\begin{tabular}{@{}l r r r r r r r r@{}}
	& \multicolumn{4}{c}{\textsc{GitHub}} & \multicolumn{4}{c}{\textsc{Yelp}}\\
	\cmidrule(l){2-5}\cmidrule(l){6-9}
           & mean & std & min &   max &  mean &   std & min &   max\\
\midrule                              
Query Size & 109  & 43 & 7   & 1,425 &    66 &    30 &   5 &   229\\
Width      &  23  &  8  & 2   &    53 &    11 &     3 &   1 &    21\\
Nesting    &   3  &  1.4  & 1   &    9 &     3 &     0.5 &   1 &     3\\
Lists      &   47  &  23  & 0   &    503 &     74 &     53 &   0 &    370\\
Response   &  79  & 67  & 0   & 2,548 & 1,301 & 2,111 &   0 & 7,363\\
\bottomrule
\end{tabular}}
\end{table}

\paragraph{Training}
The final model is selected using $5$-fold cross validation~~\cite{DBLP:conf/ijcai/Kohavi95} and the Hyperopt-sklearn optimizer~\cite{hyperopt-sklearn}.
We let our optimizer run for sixty hours for each trained pipeline for both the Yelp and GitHub datasets, exploring a total of $1,500$ combinations of models and hyperparameters, whichever of the two finishes first.
Once operators and hyperparameters are chosen, training a given model is relatively fast.

\paragraph{Results}
It is important to emphasize that finding the optimal ML pipeline was not our goal.
The main goal was to provide a framework that could be easily adapted for other APIs and datasets.
However, we need to underline the fact that, in general, the preferred estimator chosen by the Hyperopt optimizer in most of the training pipelines was the gradient boosting regressor for both GitHub and Yelp datasets.
The specific pipelines for both datasets are presented in Table~\ref{tab:piplines}.
\begin{table*}
  \begin{tabular}{l c@{$\,\rightarrow\,$}c@{$\,\rightarrow\,$}c c@{$\,\rightarrow\,$}c@{$\,\rightarrow\,$}c c@{$\,\rightarrow\,$}c@{$\,\rightarrow\,$}c c@{$\,\rightarrow\,$}c@{$\,\rightarrow\,$}c}
                    & \multicolumn{3}{c}{$m_f$} & \multicolumn{3}{c}{$m_g$} & \multicolumn{3}{c}{$m_s$} & \multicolumn{3}{c}{$m_\mathit{final}$} \\
\cmidrule(l){2-4}\cmidrule(l){5-7}\cmidrule(l){8-10}\cmidrule(l){11-13}
    \textsc{GitHub} & \textsf{SS}&\textsf{No-Op}&\textsf{GBR} & \textsf{SS}&\textsf{Poly}&\textsf{GBR}  & \textsf{SS}&\textsf{Poly}&\textsf{GBR} & \textsf{SS}&\textsf{Poly}&\textsf{GBR}  \\
    \textsc{Yelp}   & \textsf{SS}&\textsf{No-Op}&\textsf{GBR} & \textsf{SS}&\textsf{Poly}&\textsf{LR} & \textsf{SS}&\textsf{No-Op}&\textsf{GBR} & \textsf{SS}&\textsf{Poly}&\textsf{GBR} \\
    \bottomrule
  \end{tabular}
\vspace{0.5em}
\caption{Pipeline configurations for the ML models.}
\label{tab:piplines}
\end{table*}

\subsection{RQ1: Accuracy}

\begin{figure}
	\centering
	{\tabcolsep0pt
		\begin{tabular}{@{}c c@{}}
			\subfigure[GitHub static analysis.]
			{
				\includegraphics[width = 0.5\linewidth]{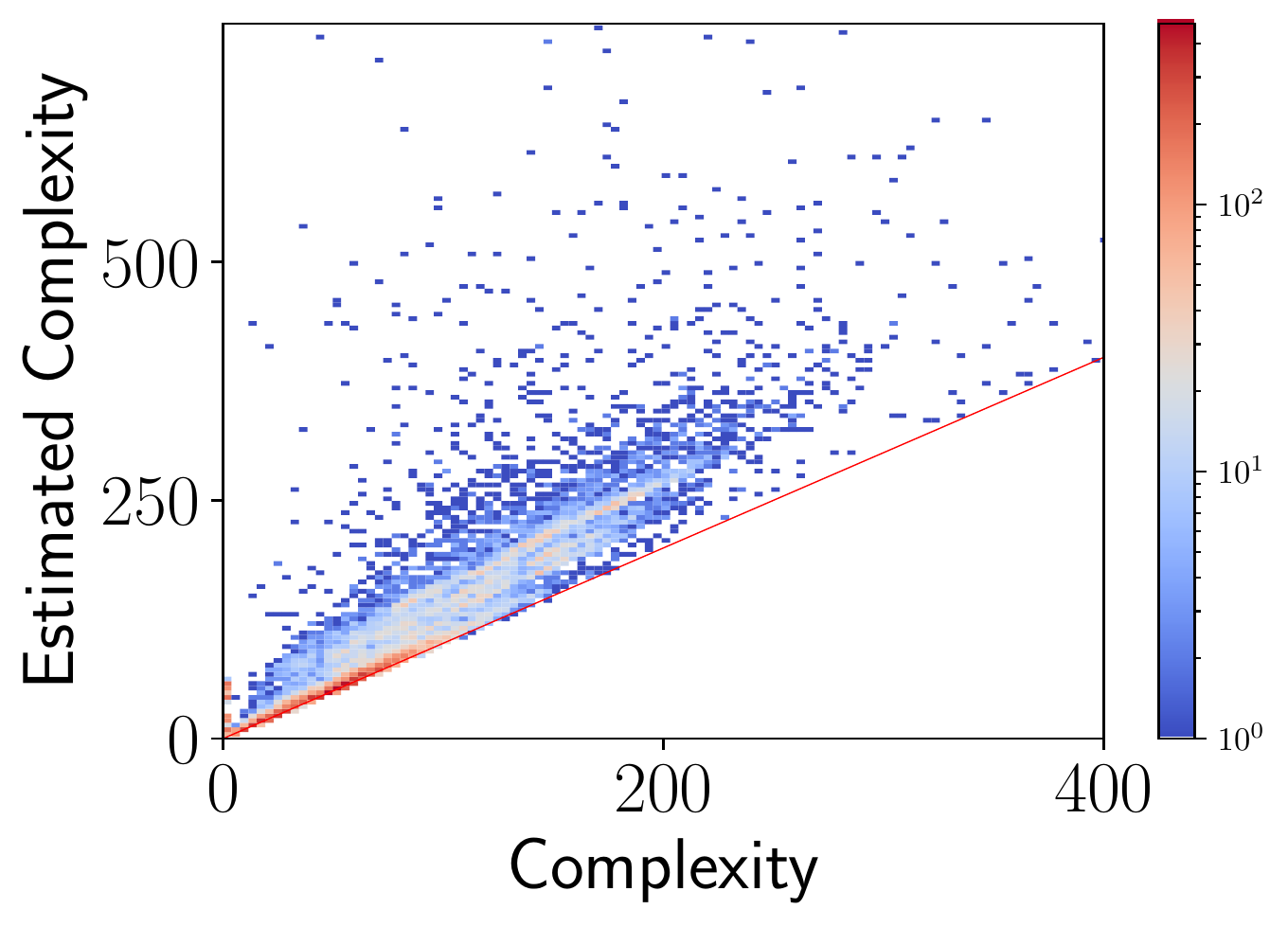}
				\label{fig:finalGS}
			}
			&\subfigure[GitHub ML estimation.]
			{
				\includegraphics[width = 0.5\linewidth]{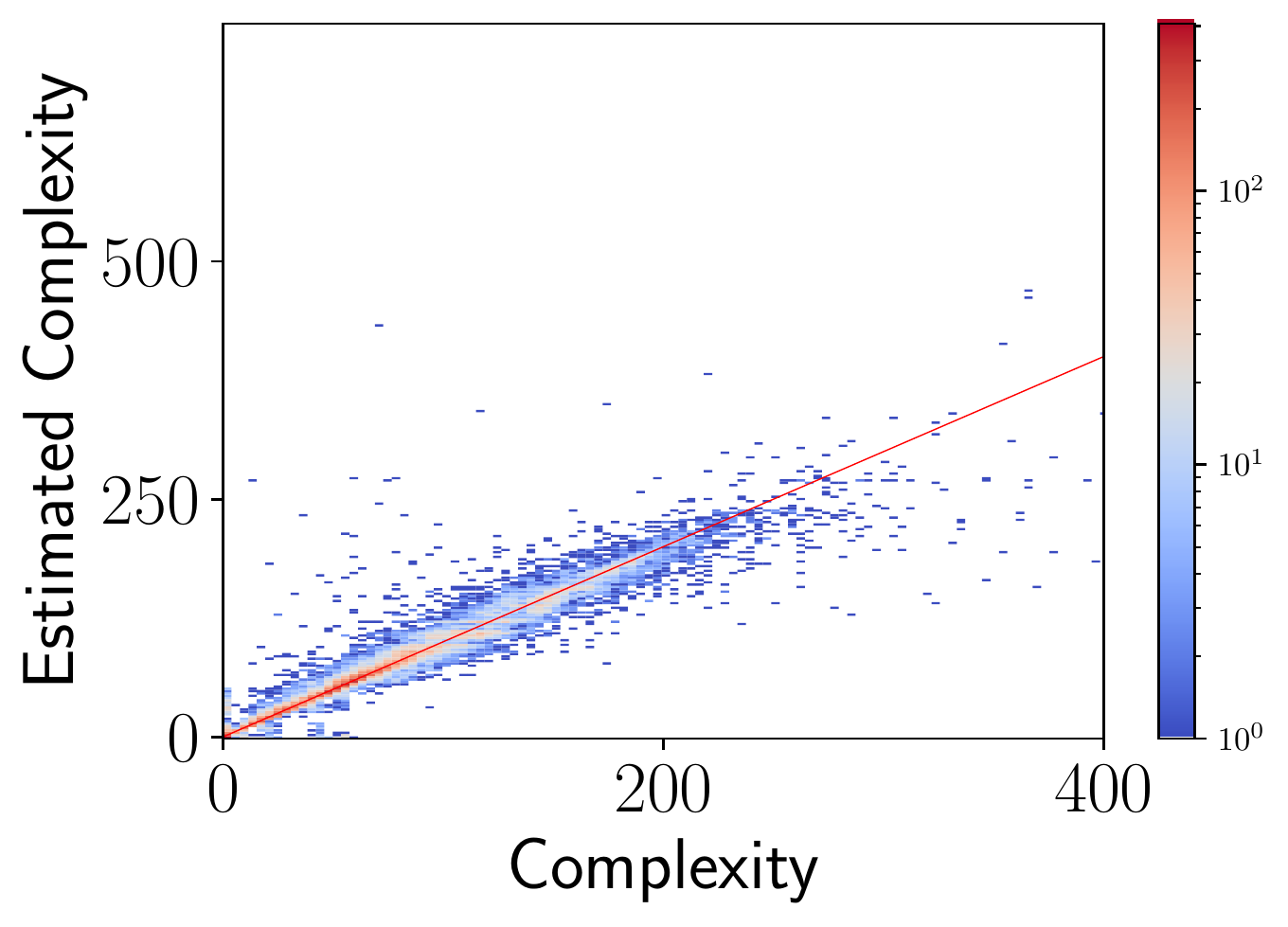}
				\label{fig:finalG}
			}
			\\
			\subfigure[Yelp static analysis.]
			{
				\includegraphics[width = 0.5\linewidth]{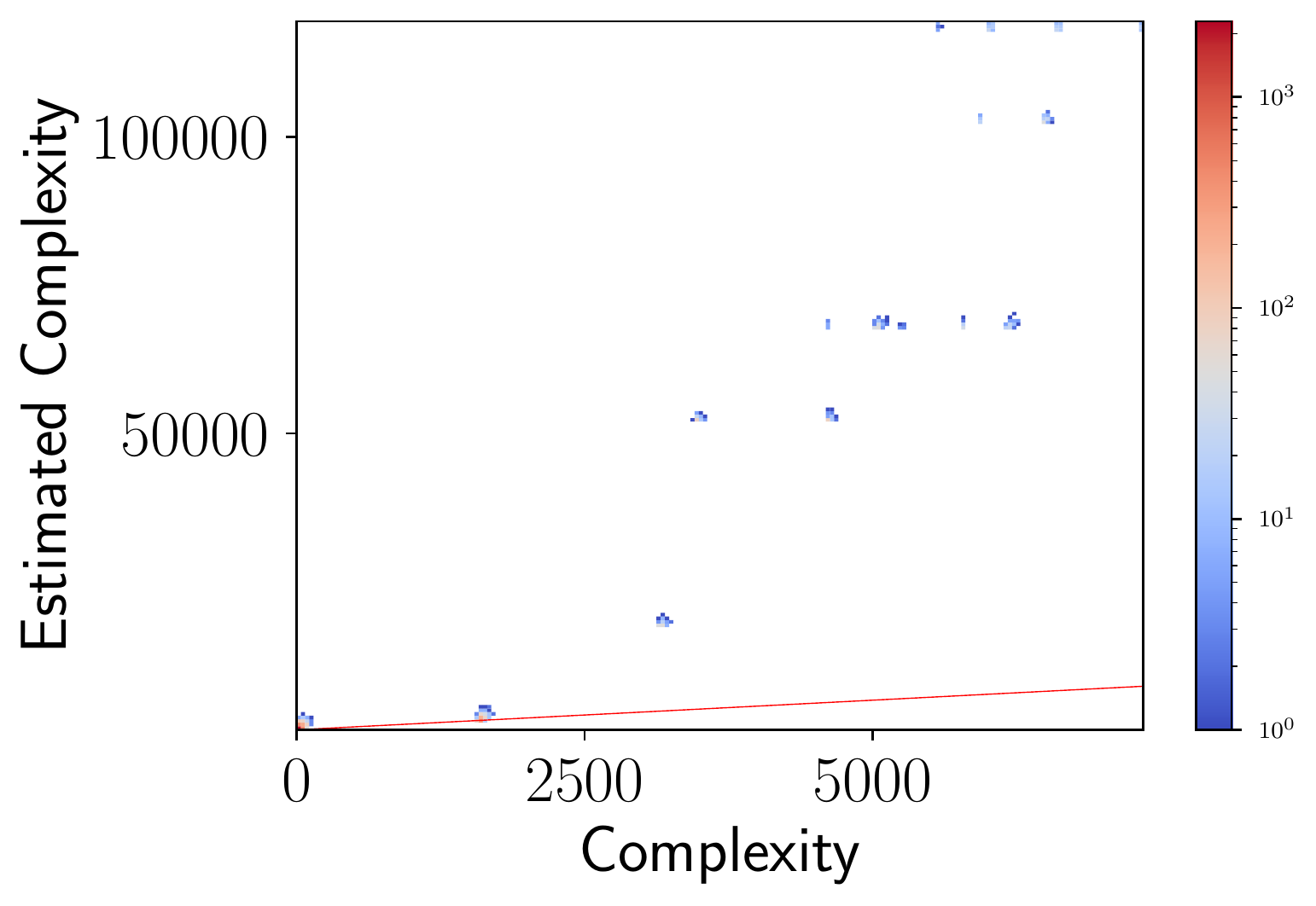}
				\label{fig:finalYS}
			}
			&\subfigure[Yelp ML estimation.]
			{
				\includegraphics[width = 0.5\linewidth]{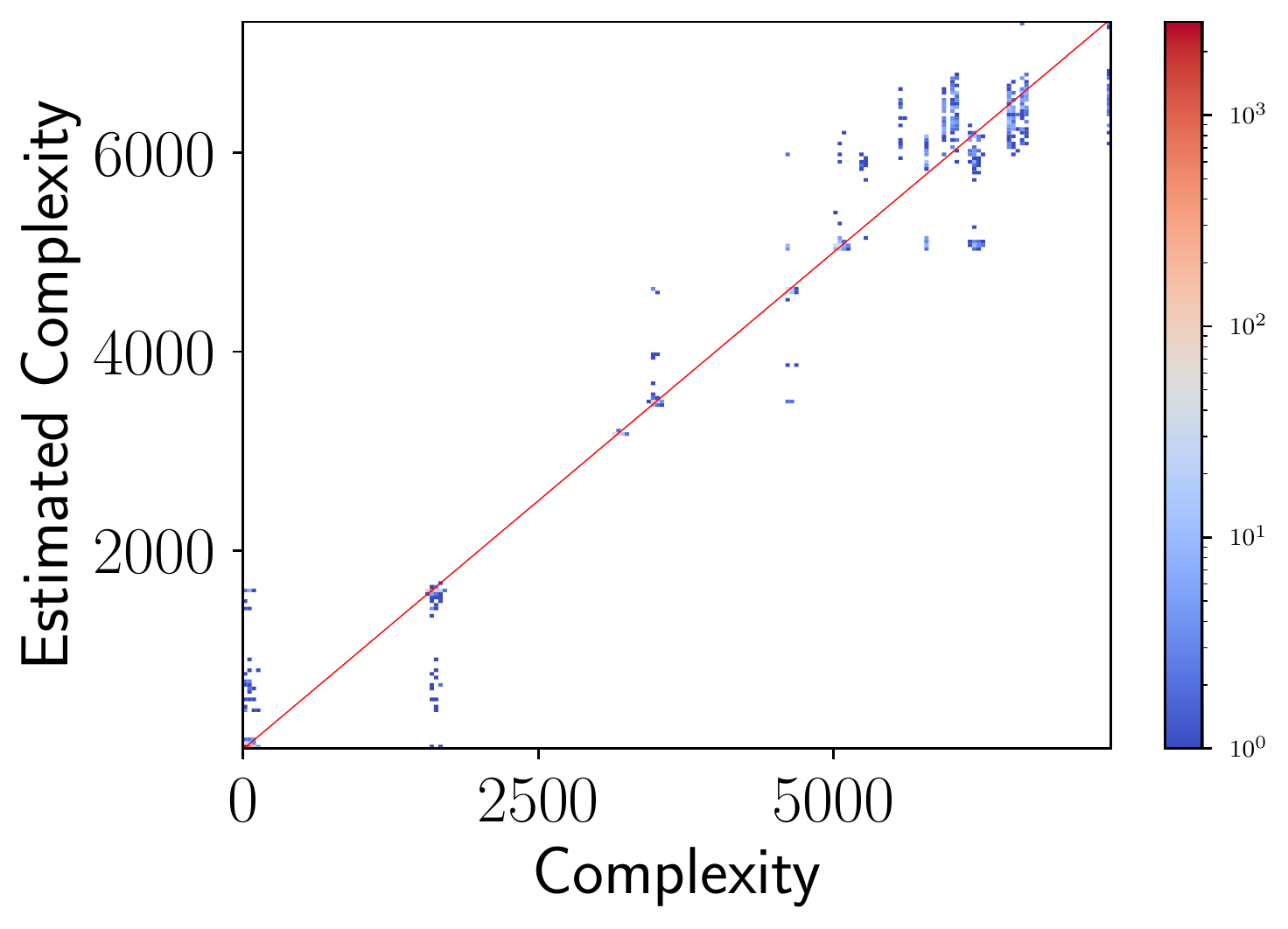}
				\label{fig:finalY}
			}
	\end{tabular}}\vspace*{-7pt}
	\caption{Comparison of complexity estimations between static analysis (left) and ML approach (right) on the GitHub (top) and Yelp (bottom) datasets. The red line represent the actual complexity of the responses and the blue dots show the estimates. Opacity indicates density.}
	\label{fig: accuracy comparison}
\end{figure}

We compare our approach to the static analysis proposed in~\cite{cha2020principled}, which was shown to outperform the three most popular libraries for computing GraphQL query cost.
Figure~\ref{fig: accuracy comparison} compares the estimations of the static analysis with those from our ML approach.
We observe that, while the static analysis guarantees an upper bound, the price in terms of over-estimation can be significant, especially with larger query sizes.
On the other hand, for both datasets, the ML estimates stay remarkably close to the actual response complexity even for the largest queries.

Note that, in order to create the visualizations for the static analysis estimates versus the ML ones, we only show the $99.8\%$ of the estimates for GitHub data and $99.5\%$ of  Yelp data. To get these percentages we first sort the static analysis estimates from the lowest to the highest and get the size corresponding to the percentage mentioned above. The reason for this is that some huge outliers dominated the graphs, impeding the reader from extracting any meaningful conclusions from them. 

These plots also illustrate the difference between the two APIs.
The random query generator is able to smoothly explore the complexity space of the GitHub API.
For the Yelp API, however, queries form dense clusters that are distant from each other, resulting in the 
static analysis' estimates precision degrading significantly as the query complexity increases.

\begin{table}
	\centering 
	\begin{tabular}{@{}l r r r r r@{}}
		& \multicolumn{2}{c}{\textsc{Static}} &\hspace{2em}& \multicolumn{2}{c}{\textsc{ML}}\\
		\cmidrule(l){2-3}\cmidrule(l){5-6}
		& MAE & std && MAE & std\\
		\midrule
		\textsc{GitHub} &31.5 & 263.8 && 8.2 & 35.5  \\	
		\textsc{Yelp} &14,180.5 & 30,827.9 && 60.7 & 180.4  \\	
		\bottomrule 
	\end{tabular} 
	\vspace{0.5em}
	\caption{MAE Comparison between the ML approach and the static analysis.}
	\label{table:accuracy comparison}
\end{table}

To quantify the precision of the ML approach compared to the static analysis, we computed the Mean Absolute Error (MAE) between the estimated cost and the actual cost of the responses on the two datasets. Results are summarized in Table~\ref{table:accuracy comparison}.
$$
\textrm{MAE} = 1/n \sum_{i=1}^n |c_i-\hat{c}_i|
$$

\begin{figure}
	\centering
	{\tabcolsep0pt
		\begin{tabular}{@{}c c@{}}
			\subfigure[GitHub error distributions.]
			{
				\includegraphics[width = 0.5\linewidth]{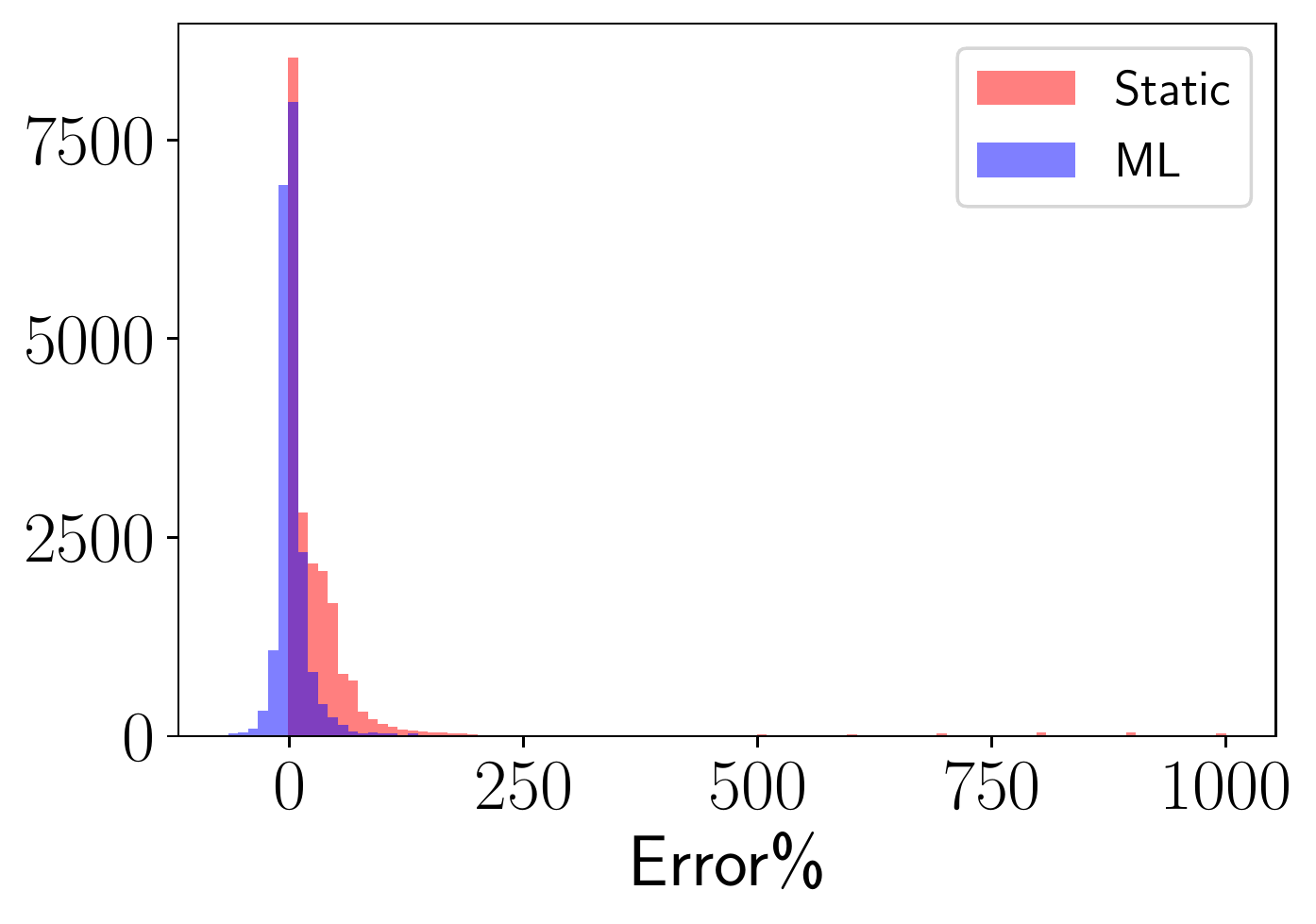}
				\label{fig:errorG}
			}
			&\subfigure[Yelp error distributions.]
			{
				\includegraphics[width = 0.49\linewidth]{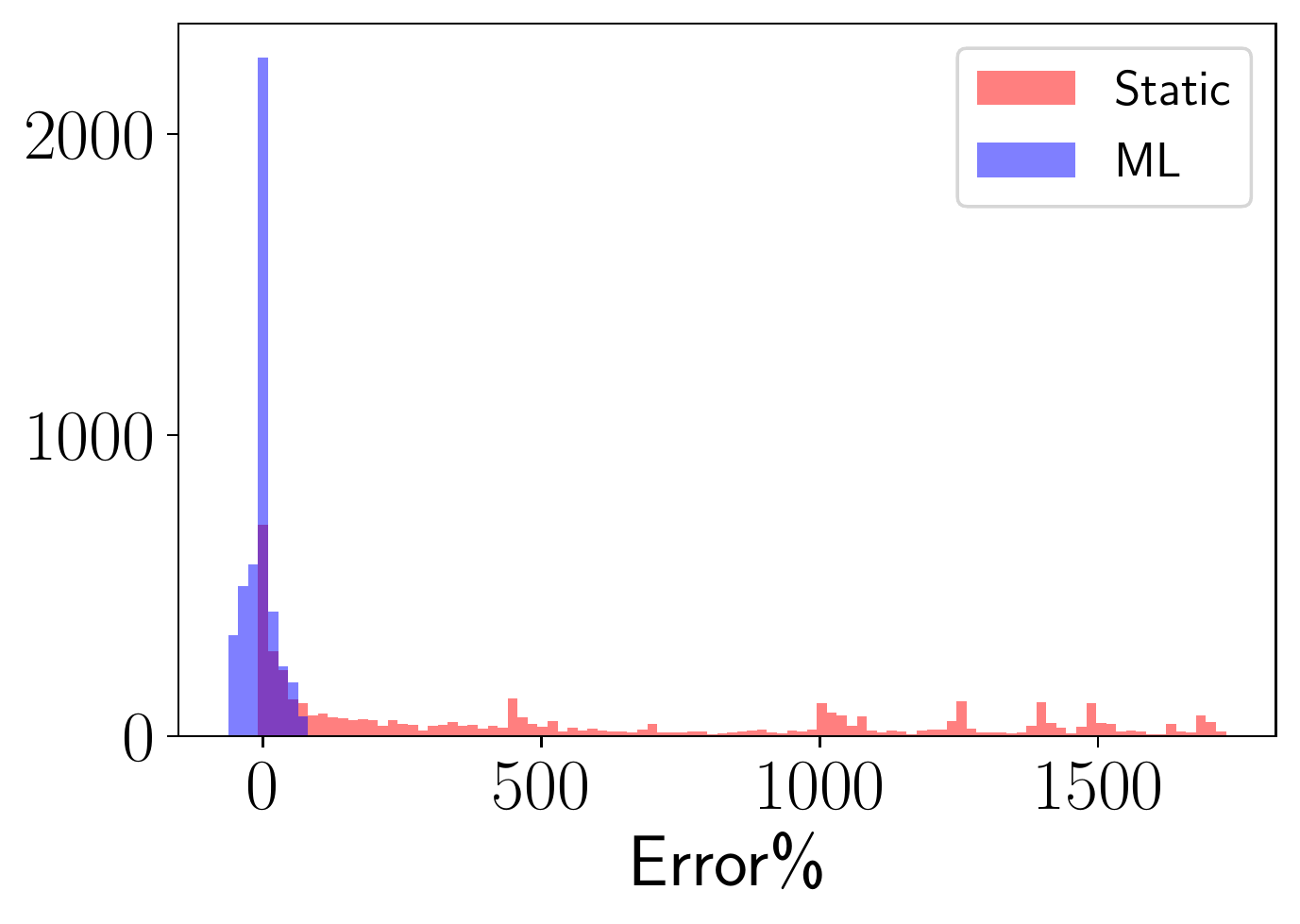}
				\label{fig:errorY}
			}
		
	\end{tabular}}\vspace*{-7pt}
	\caption{Error percentage distribution for the ML and static predictions. For visual purposes we have removed outliers with error\% reaching up to $120,000$\% in static analysis.}
	\label{fig:error_dist}
\end{figure}

\noindent
For both datasets, the accuracy gain of the ML approach compared to the static analysis is striking both in terms of average value, and standard deviation.
This further validates the observation that ML approach is accurate for large queries, which are challenging for the static analysis. To further underline our claim, we present in \Cref{fig:error_dist} the raw error distribution percentages of the ML and static analysis. Given a query with response cost~$c$ and a prediction~$\hat{c}$, we define the error percentage as $Error\% = \frac{\hat{c}-c}{c}$.

\subsection{RQ2: Features Selection}

As described in Sections~\ref{sec:features} and~\ref{sec:learning},
the ML estimates are based on three groups of features, namely summary features (including the result of the static analysis), field features, and graph embedding features. But are all these features necessary?
To answer this question, we looked at
estimates obtained using each group of features separately.
Table~\ref{table:features comparison} summarizes the results.
We observe that for both datasets, none of the feature alone is competitive with the stacked ensemble presented in Figure~\ref{fig:process},
that combines the results of cost estimation models trained from all
three groups of features separately.

The performance of each group of features depends on the dataset.
For instance, while the \emph{summary features} give reasonable estimates for both datasets, the \emph{field features} are much more useful for GitHub than for Yelp. 
This could be related to the underlying structure of the two datasets as well as to the data generation process.
Table~\ref{table:features comparison} also shows that the automatic feature extraction of the neural networks used to build the \emph{graph embedding} features fails to produce accurate estimates for either dataset, underscoring the importance of the more descriptive features.

\begin{table}
\caption{\label{table:features comparison}Accuracy comparison for each feature}
\centerline{\begin{tabular}{@{}l r r r r r@{}}
   & \multicolumn{2}{c}{\textsc{GitHub}} &\hspace{0.5em} & \multicolumn{2}{c}{\textsc{Yelp}}\\
  \cmidrule(l){2-3}\cmidrule(l){5-6}
                     &  MAE &   std  &&     MAE &     std\\
  \midrule 
  Summary features   &  8.7 &  36.4  &&   102.4 &   280.8\\
  Field features     & 14.9 &  40.2  &&   320.8 &   715.6\\
  Embedding features & 31.58 &  45.7  && 880.9 & 813.4\\
  Final combination  &  8.2 &  35.5  &&   60.7 &   180.4\\
  \midrule
  Static analysis    & 31.5 & 263.8 &&  14,180.5 & 30,827.9\\
  \bottomrule 
\end{tabular}}
\end{table}

To delve a bit deeper into the feature analysis, we performed an independent univariate test to capture the relationship between the features and the target~(here the query cost). We want to understand how each feature, if used separately, impact the target values.
For this experiment to be meaningful, we used the summary features ($\mathcal{F}_s$) and the field features ($\mathcal{F}_f$), which are interpretable, as they are related to known quantities.
Given a feature $X_i$ and its corresponding data $x_{i1}, \ldots, x_{iN}$ and the target query costs $Y = \{y_1, \ldots, y_N\}$, we calculate the mutual information
between $X_i \in \mathcal{F}_s \cup \mathcal{F}_f,$ and $Y$ using the scikit-learn's implementation of the mutual information metric.\footnote{\url{https://scikit-learn.org/stable/modules/generated/sklearn.feature_selection.mutual_info_regression.html}}
This metric tries to capture the dependence between two random variables.
We report the top-$10$ features with the biggest dependencies~(with respect to this metric) to the target in \Cref{tab:topk}.

It is noticeable that a subset of the summary features, which are common in both datasets, can be found in the top positions for both datasets. This could provide some further insight on why the model $m_s$ constructed by these features performs better in comparison to the models $m_f$, and $m_g$ as we see on table \ref{table:features comparison}.
\begin{table}
	\centering
	\begin{tabular}{l l}
		\textsc{GitHub} &\textsc{Yelp}\\
		\toprule
		$\mathcal{F}_s$: \textbf{Lists}       & $\mathcal{F}_s$: ResolveComplexity\\
		$\mathcal{F}_f$: MarketPlaceCategories& $\mathcal{F}_s$: \textbf{TypeComplexity}\\
		$\mathcal{F}_f$: Licenses             & $\mathcal{F}_f$: Code\\
		$\mathcal{F}_f$: Permissions          & $\mathcal{F}_s$: \textbf{Lists}\\
		$\mathcal{F}_s$: \textbf{Nesting}     & $\mathcal{F}_s$: \textbf{Nodes}\\
		$\mathcal{F}_f$: Label                & $\mathcal{F}_f$: Parent Categories\\
		$\mathcal{F}_f$: Description          & $\mathcal{F}_f$: Country Whitelist \\
		$\mathcal{F}_s$: \textbf{TypeComplexity} & $\mathcal{F}_f$: Country Blacklist \\
		$\mathcal{F}_f$: \textbf{Sum of Variables}        & $\mathcal{F}_s$: \textbf{Nesting} \\
		$\mathcal{F}_s$: \textbf{Nodes}       & $\mathcal{F}_s$: \textbf{Sum of Variables} \\
		\bottomrule	
	\end{tabular}
        \vspace{0.5em}
	\caption{Top-$10$ features with the largest dependency to the query costs based on the mutual information criterion for GitHub and Yelp. In bold text we see the common summary features
	in top positions for both datasets.}
	\label{tab:topk}
\end{table}

\subsection{RQ3: Practicality}
\label{sec:practicality}

Now that we have access to accurate complexity estimates, the main question is: \emph{how useful are these estimates in practice?}
API managers offer, through a client-selected plan, a rate limit, allowing a certain number of points per time window. Points could be attributed to individual REST calls or to the
query cost in the case of GraphQL. Several API management vendors implement these strategies~\cite{APIConnect, APIgee, 3Scale}.
To mimic this behavior, we built a simulator that acts as an API manager whose goal is to filter queries based on the client plan. We 
 select a threshold of points to represent the rate limit on a given time window.

First, the client sets a threshold, that is, the maximal aggregate cost that the client is willing to pay for a query.
Then the simulator acts as a gateway between the API and the client, rejecting queries for which the estimated cost is above the threshold.
To evaluate the benefit of our approach, we compare the acceptance rate of a simulator relying on the static analysis against the acceptance rate of a simulator relying on our ML approach.

\begin{figure}
	\centering
	{\tabcolsep0pt
		\begin{tabular}{c c }
			\subfigure[GitHub]
			{
				
				\includegraphics[width = 0.49\linewidth]{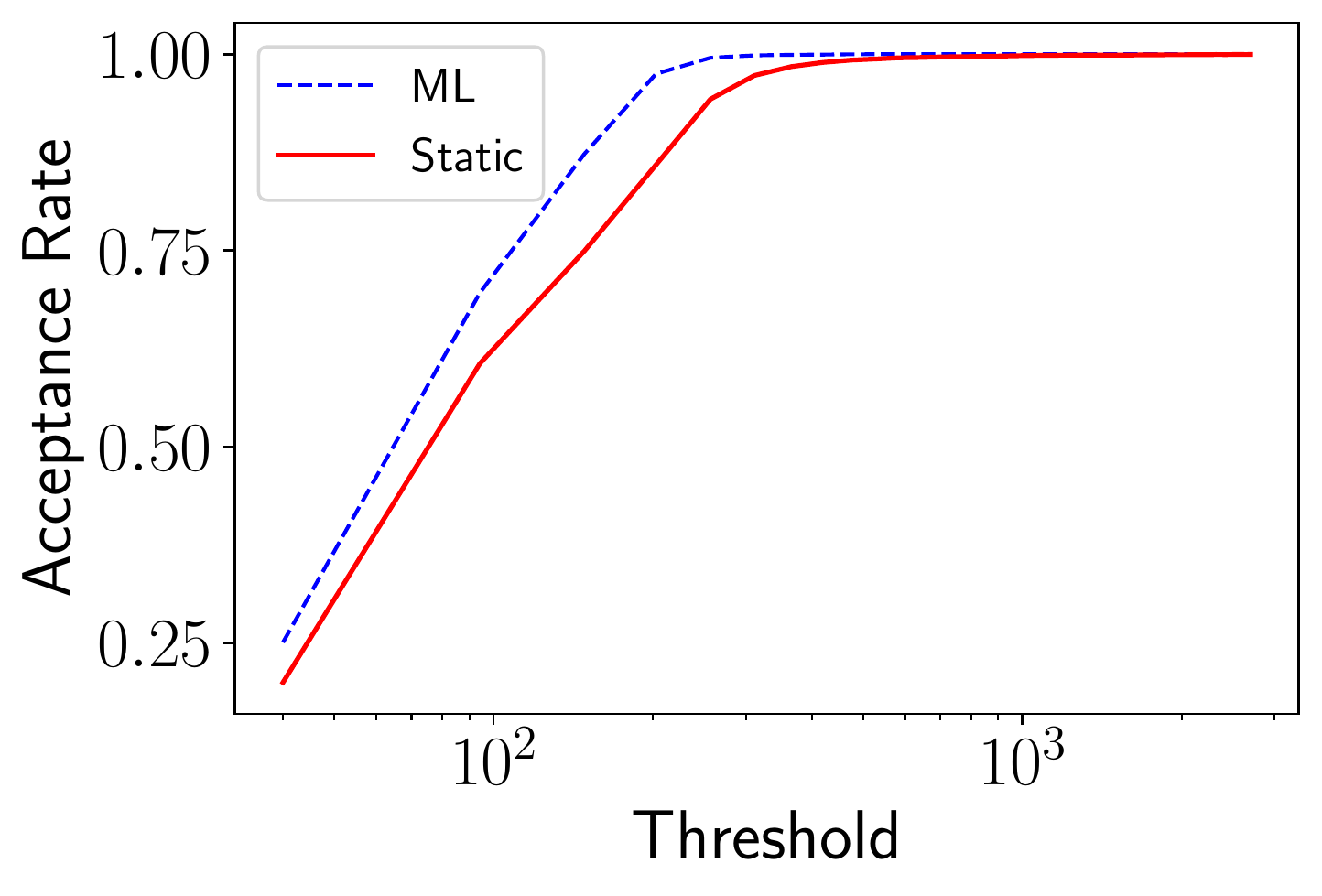}
				\label{fig:embY}
				
			}
			&\subfigure[Yelp]
			{
				\includegraphics[width = 0.49\linewidth]{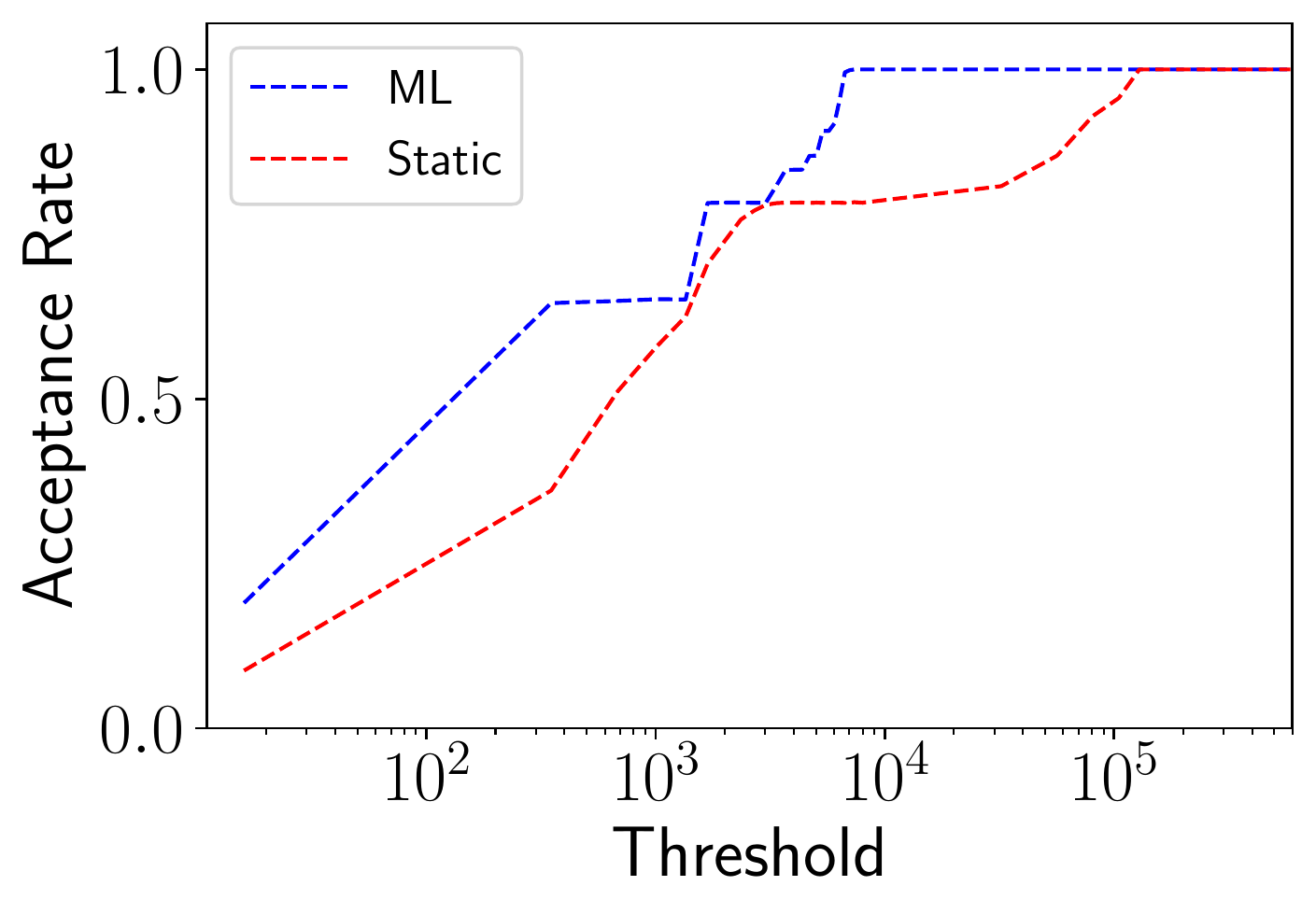}
				\label{fig:fieldY}
			}
		\end{tabular}}
		\caption{Comparison of the acceptance rate against increasing threshold for the static analysis (red) and the ML approach (blue). The acceptance rate is computed as the number of queries whose sum of estimated complexity is below the threshold level divided by the number of sampled queries.}
		\label{fig:acceptance rate}
\end{figure}

Figure~\ref{fig:acceptance rate} shows the evolution of the acceptance rate for increasing values of the simulation threshold. We used a different range of thresholds for the experiments in Yelp and GitHub respectively due to their specific characteristics~(in general Yelp contains queries with larger costs).
The results are averaged over 1,000 simulations, and for each simulation we randomly select 1,000 queries.
Overall, as expected, the ML cost estimation policy is able to accept a bigger proportion of queries for both APIs.
The staircase shape of the Yelp results can be explained by the peculiar cluster-like distribution of query complexity in the dataset (see Figure~\ref{fig: accuracy comparison}).
For the same reason, the static analysis plateaus at $80\%$ until the threshold is considerably larger due to the substantial overestimation within this range. When the threshold reaches a high enough value, the static analysis reaches $100\%$ acceptance rate too.

\begin{figure}
	\centering
	{\tabcolsep0pt
		\begin{tabular}{c c }
			\subfigure[GitHub, Threshold: $44$]
			{
				\includegraphics[width = 0.49\linewidth]{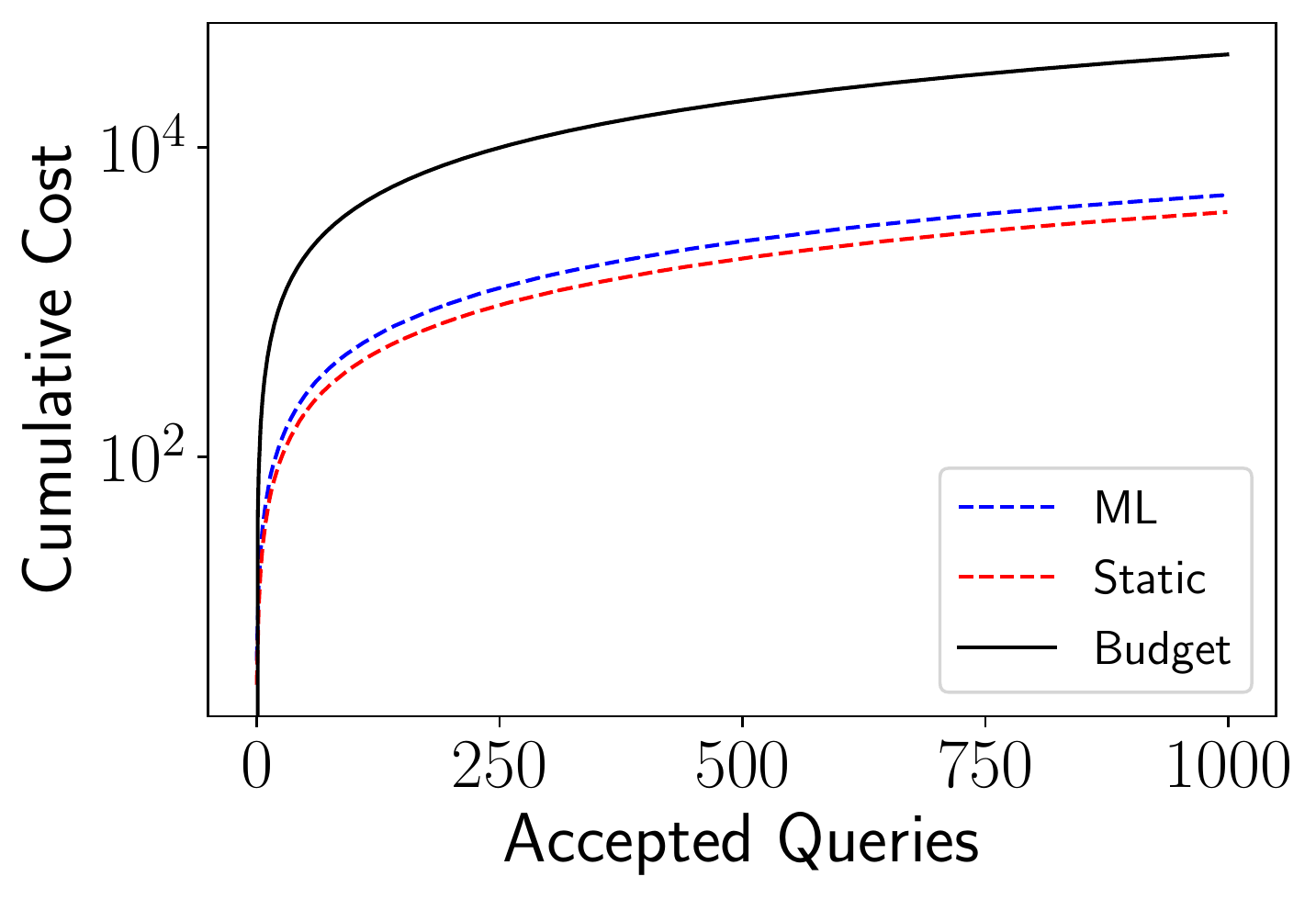}
				\label{fig:th0}
			}
&
			\subfigure[GitHub, Threshold: $474$]
			{
				\includegraphics[width = 0.49\linewidth]{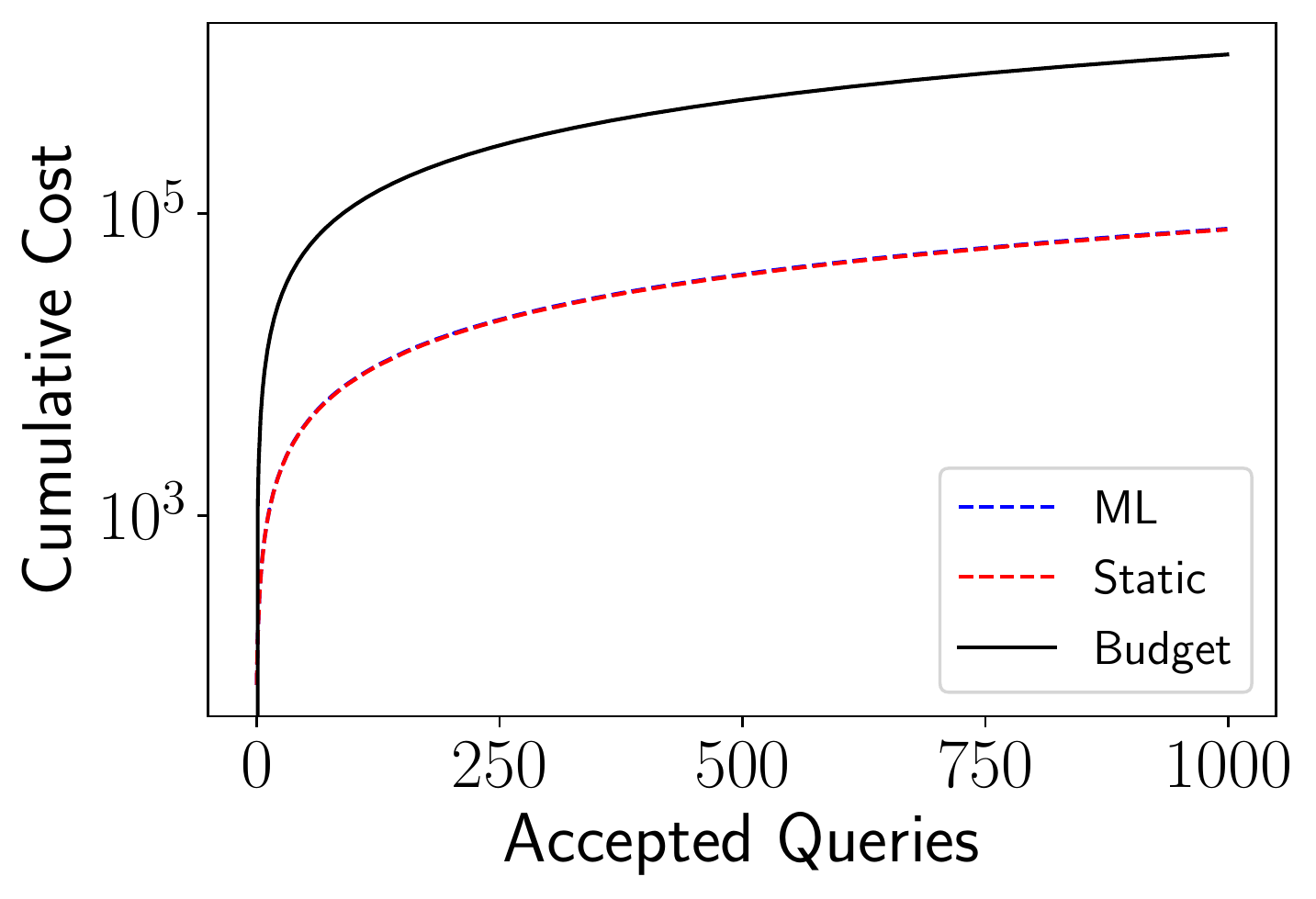}
				\label{fig:th2}
			}
			\\
                        \subfigure[Yelp, Threshold: $15$]
			{
				\includegraphics[width = 0.49\linewidth]{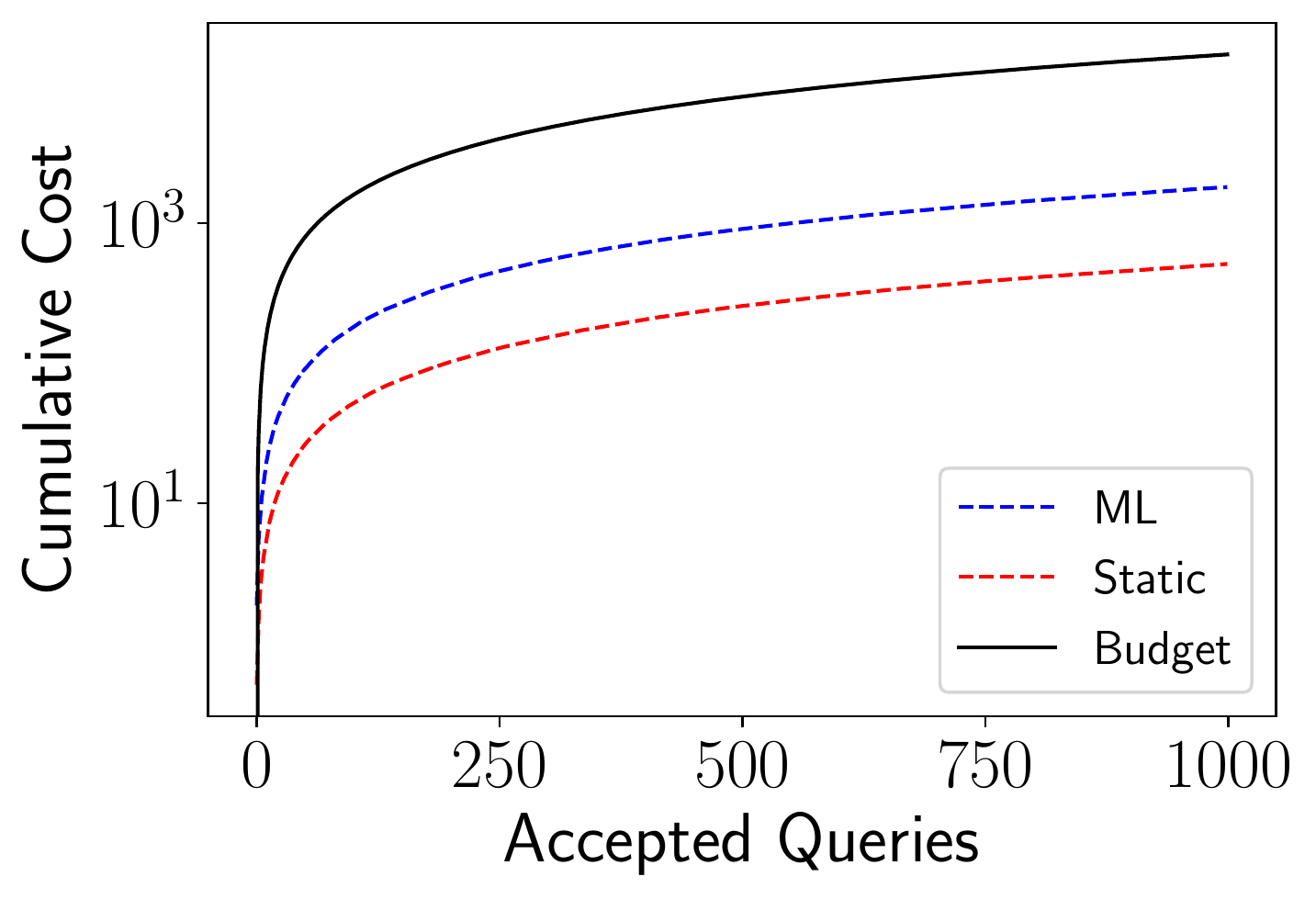}
				\label{fig:th0Y}
			}
&\subfigure[Yelp, Threshold: $6710$]
			{
			\includegraphics[width = 0.49\linewidth]{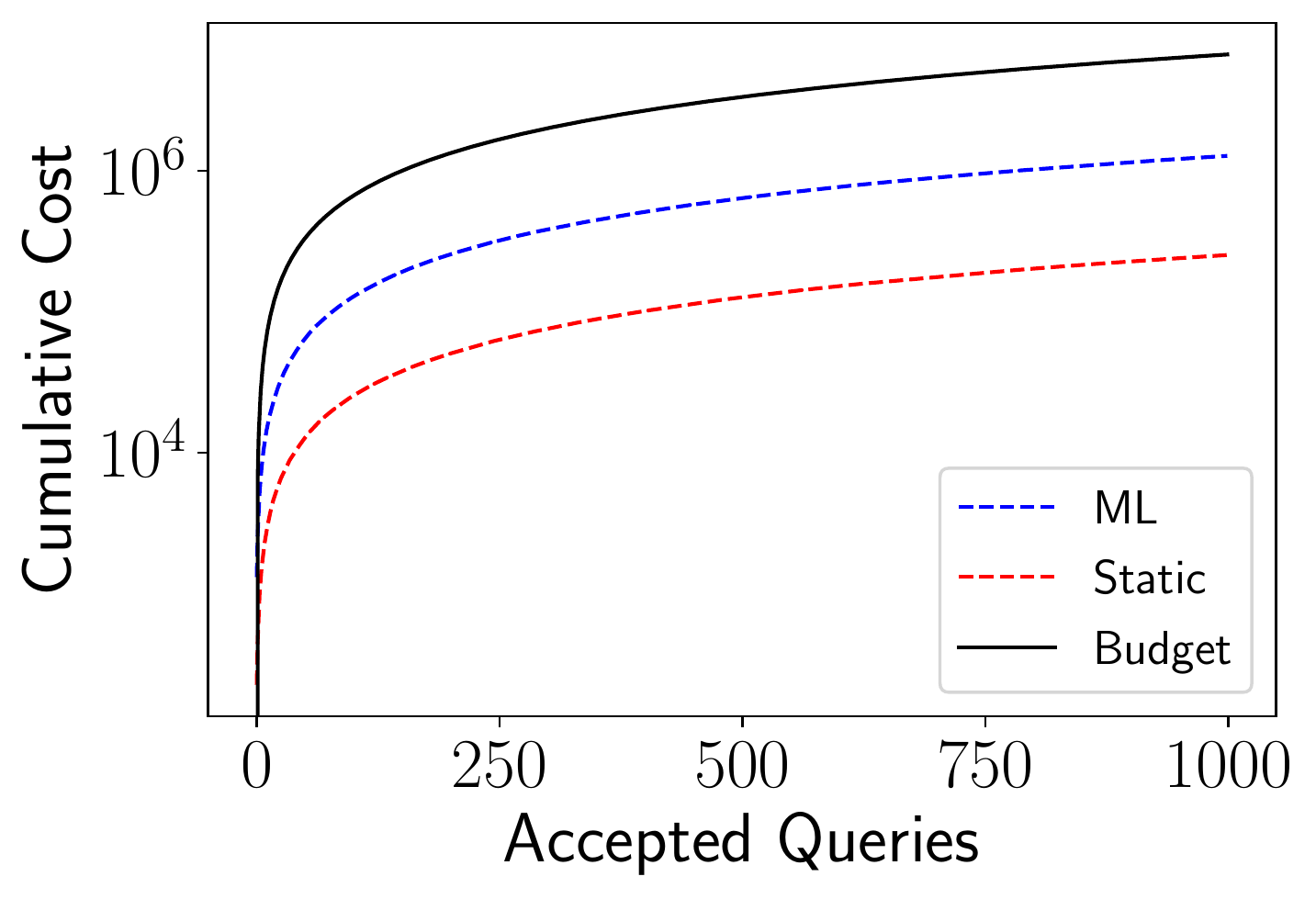}
			\label{fig:th2Y}
			}
	\end{tabular}}\vspace*{-7pt}
	\caption{Cumulative cost incurred by the the static analysis and ML approach on accepted queries.
          The \textsf{budget curve} is the cumulative sum of the budget allocated per query. The budget of each query is considered constant and given by the value of the threshold.
         The \textsf{static curve} and the \textsf{ML curve} show the cumulative sum of the response complexity of the queries that were accepted because their expected complexity computed respectively by the static analysis and ML approach was below the threshold.}
	\label{fig:sims}
\end{figure}

Figure~\ref{fig:sims} shows the results of the simulation for a series of threshold values.
The low threshold is such that the complexity of $25\%$ of the queries in the dataset is under the threshold (25th percentile). The high threshold is five times the 75th percentile.
While it is still possible that a few queries go over budget due to the approximate nature of the ML estimates, the average cumulative cost (red) always stays below the maximum budget (black).
This suggests that our approach can be used to implement rate limiting policies.

For the smallest threshold value, the simulator can only accept very few queries for which the difference with the maximum possible cost can be significant.
The cumulative cost curve thus remains clearly under the budget line and there is no noticeable difference between the static analysis and the ML estimates.
For the largest threshold values in general,  the ML estimates curve is much closer to the budget limit than the static analysis curve which indicates a better utilization of the available budget.
For this experiment, the benefit of the ML predictor compared to the static predictor is limited for GitHub (\Cref{fig:th0,fig:th2}) but more significant for Yelp (\Cref{fig:th0Y,fig:th2Y}).

\begin{figure}
	\centering
	{\tabcolsep0pt
		\begin{tabular}{c c }
			\subfigure[GitHub, Threshold Violations]
			{
				\includegraphics[width = 0.49\linewidth]{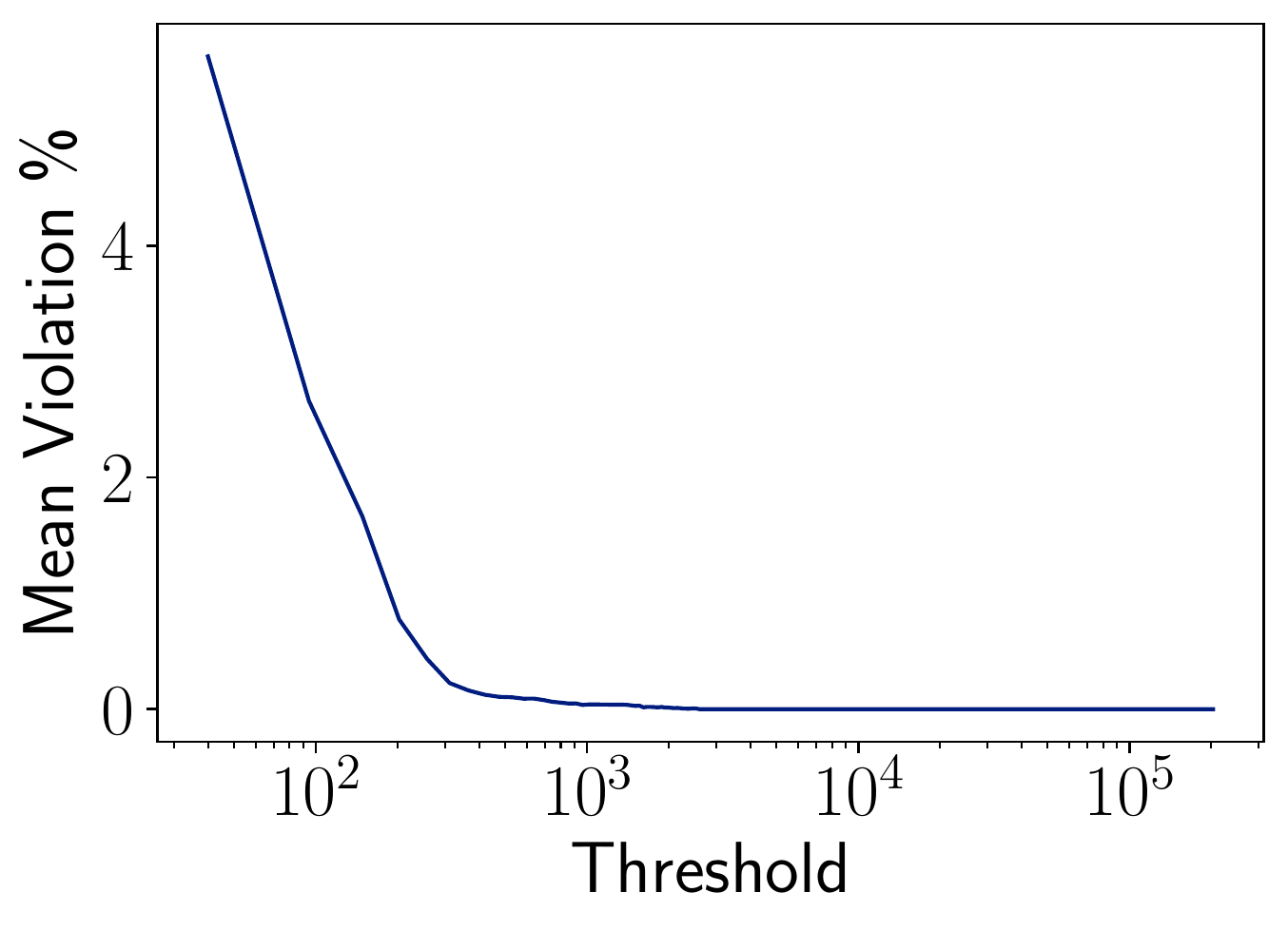}
				\label{fig:viol}
			}
			&\subfigure[GitHub, Violations Above Threshold]
			{
				\includegraphics[width = 0.49\linewidth]{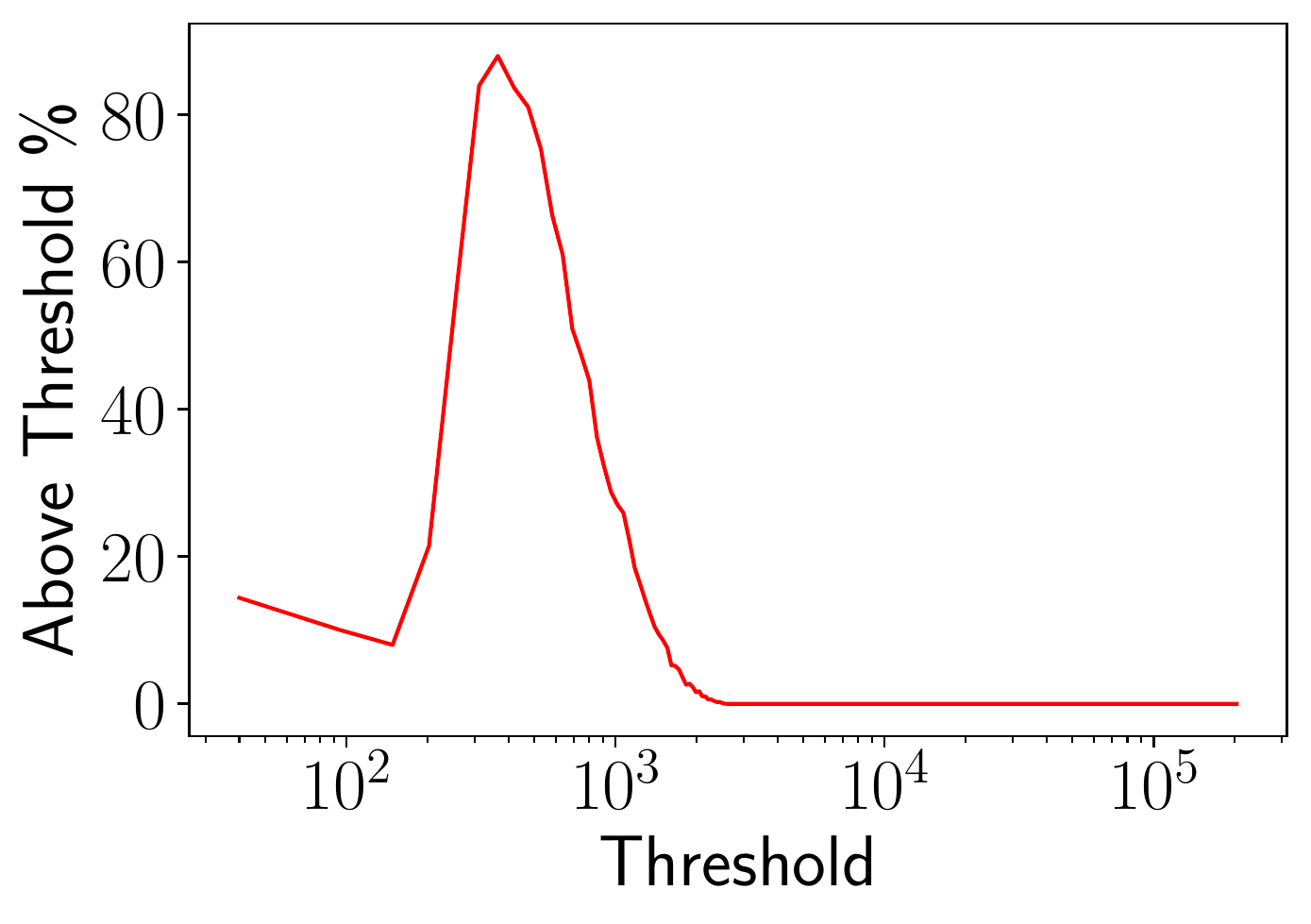}
				\label{fig:viola}
			}\\
			\subfigure[Yelp, Threshold Violations]
			{
				\includegraphics[width = 0.49\linewidth]{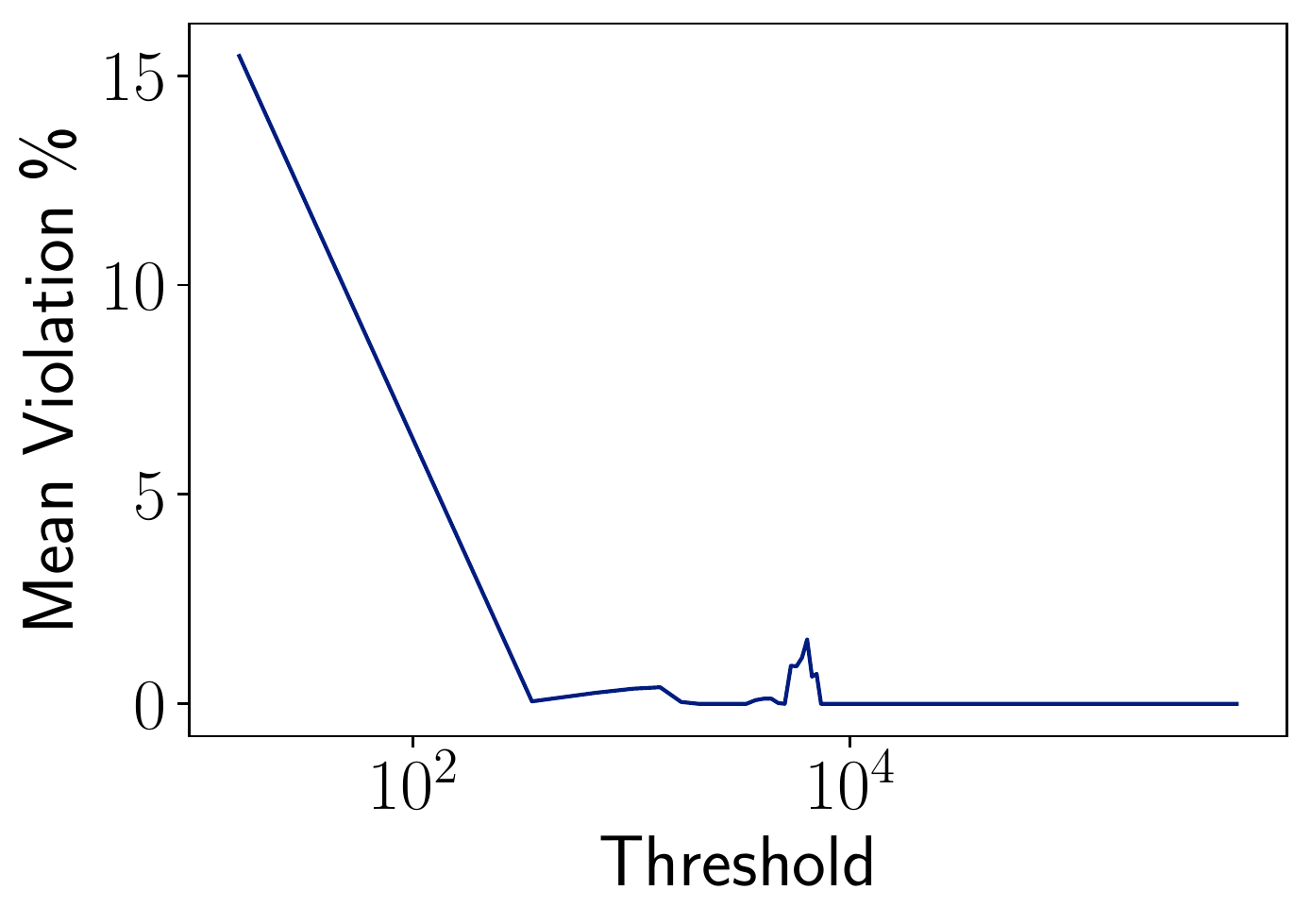}
				\label{fig:violY}
			}
			&\subfigure[Yelp, Violations Above Threshold]
			{
				\includegraphics[width = 0.49\linewidth]{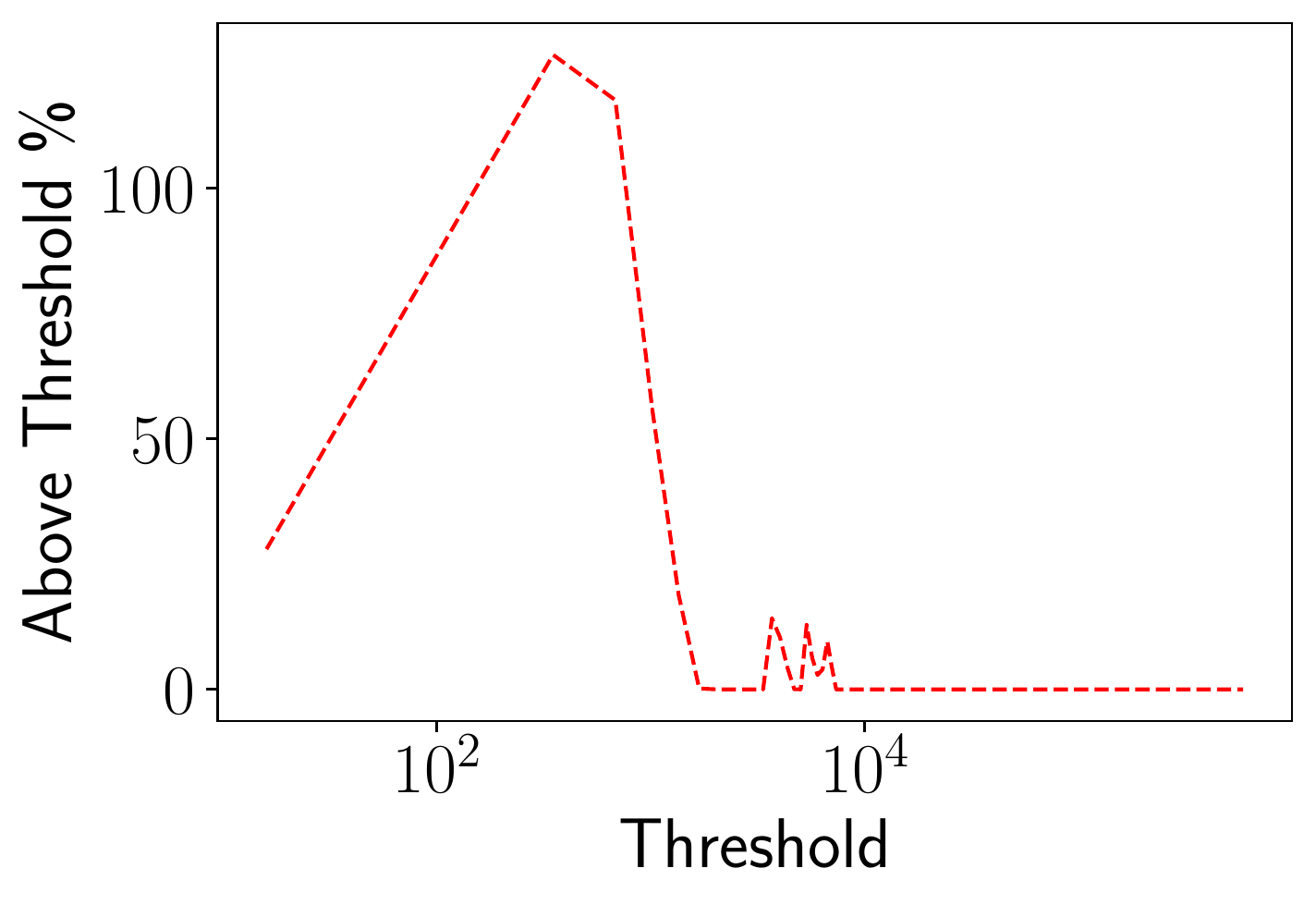}
				\label{fig:violaY}
			}
		
	\end{tabular}}\vspace*{-7pt}
	\caption{Average percentage of violations per threshold and average amount of violation percentage per threshold from the ML estimates. Figures (a) and (c) show how many accepted queries violate a given budget on average. Figures (c) and (d) show how much accepted queries violate the budget on average.}
	\label{fig:violations}
\end{figure}

Finally we conclude this experiment by exploring what is the average budget violation per threshold and when there is a violation what is the average percentage the ML predictor exceeds the budget by. Figure~\ref{fig:violations} showcases that in general there is a very small percentage of violations which quickly reaches zero as the threshold-budget increases. For GitHub, we observe that at most $6\%$ of the queries violate the budget and they are at most $80\%$ above it, see figures~\ref{fig:viol}, and~\ref{fig:viola}. For Yelp, we observe a $15\%$ of violations on small thresholds with a maximum violation of a little above $100\%$ above the threshold as figures~\ref{fig:violY}, and~\ref{fig:violaY} indicate.

\subsection{RQ4: Robustness}
\label{sec:robustness}

The ML paradigm provides accurate estimates
with high probability,
but it opens the door to the
undesirable possibility of cost underestimation.
Due to possible underestimation, using the ML prediction to filter queries can result in 
accepting a
query whose actual cost is above the budget threshold.
This is not possible with the static analysis as it calculates and upper bound cost.
The ML's under- and overestimation of costs are
generally small and statistically
fluctuate around the actual cost,
averaging out in the long run as shown in Section~\ref{sec:practicality}.
However, it paves the way to adversarial malicious queries.
The goal in this section is to evaluate the extent to which the
ML query cost estimate is robust to malicious requests.

Unfortunately, there is no single definition of a
malicious request in the context of database management and web communication.
Intuitively, a malicious query is disguised,
presenting as a query of small or minimum size, yet
producing a result that is huge or of maximum size.
Small queries can produce exponentially sized output.
For instance, \citet{cha2020principled} present an example of a
real-world pathological query in which they recursively request
the same first 100 issues from the Node repository using the
GitHub GraphQL API.
To study the robustness of the ML
paradigm against malicious queries, we set up a simple experiment that relies
on two basic properties of malicious queries,
\begin{itemize}
\item The query is disguised, presenting as a small input but nevertheless
  realizing the upper bound cost computed by the static analysis.
\item
  The actual cost and matching upper bound cost are huge.
\end{itemize}
It is the second property that makes the query malicious and capable of
breaking the system. This is the regime we would like to be robust to.

To simulate such a malicious
query, we pick a small generic query-example from the dataset as
the ``disguised'' query.
To craft a malicious query, we then assume that its upper bound cost is huge,
keeping all other features constant, and we also assume
that the actual cost matches the upper bound.
Hence, for an original query-example with a feature vector ${(x_1,\ldots,x_N,\text{static bound})}$,
we create a feature vector ${(x_1,\ldots,x_N,\text{huge static bound})}$ which corresponds to a simulated malicious query.
Note that the malicious query corresponding to the simulated feature vector may not be realizable.

We then compute the ML
predicted cost be for the simulated malicious query. If the prediction
is a small cost, then the query could be accepted and break the system. For the ML to be robust, its prediction on the
simulated malicious query should be large, ideally increasing rapidly as
the static bound increases.

\begin{figure}
	\centering
	{\tabcolsep0pt
		\begin{tabular}{c c }
			\subfigure[GitHub]
			{
				\includegraphics[width = 0.48\linewidth]{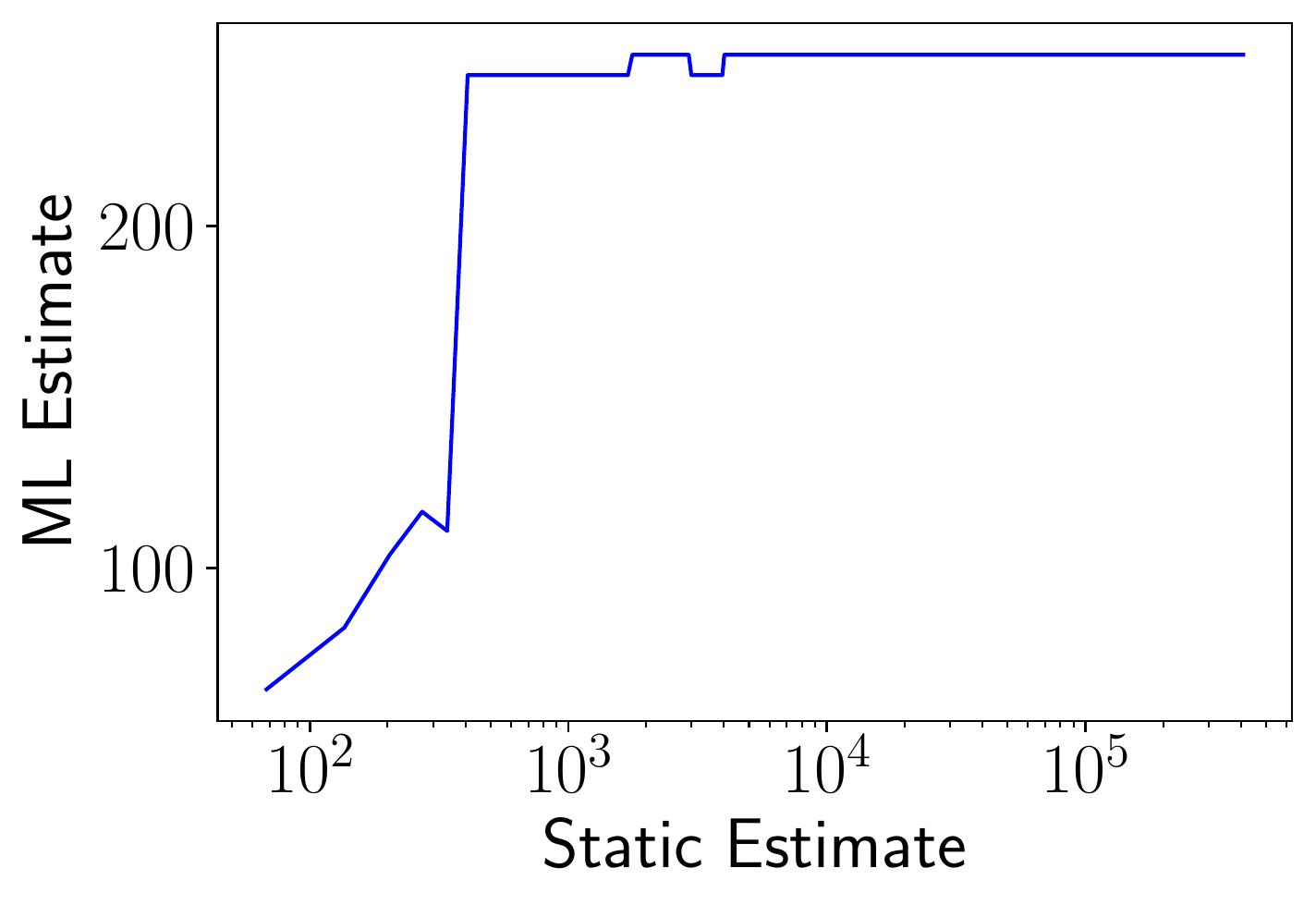}
				\label{fig:bound}
			}
			&\subfigure[Yelp]
			{	\includegraphics[width = 0.49\linewidth]{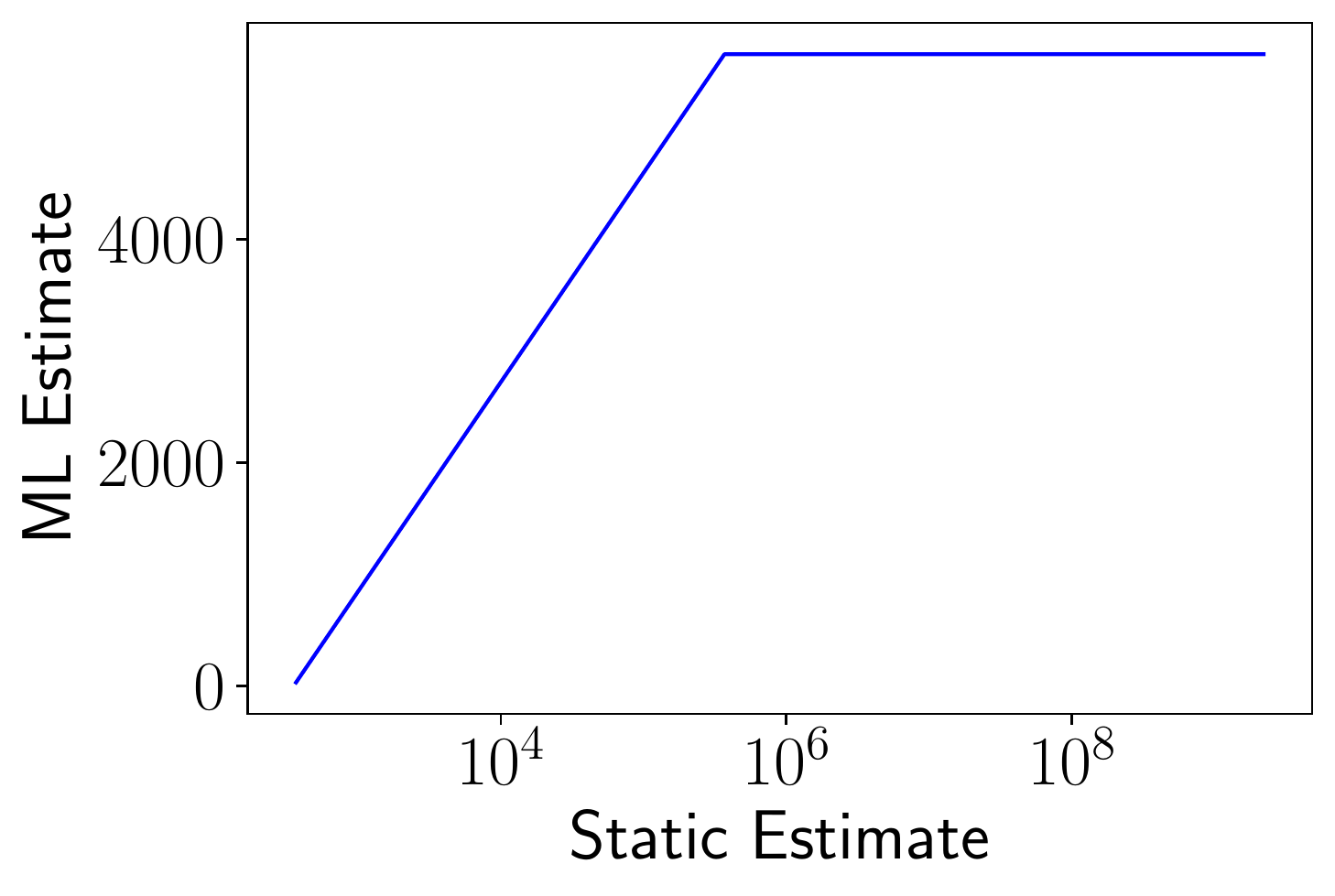}
				\label{fig:boundY}
			}
			
	\end{tabular}}\vspace*{-7pt}
	\caption{ML estimates as static analysis estimates increase.}
	\label{fig:robust}
\end{figure}

We have performed this robustness analysis on random queries selected from
the GitHub and Yelp datasets. The qualitative dependence of cost on
upper bound cost is consistent for all cases we tried, and we present a
representative result in 
Figure~\ref{fig:robust}. Notice that as the
upper bound cost
increases, the ML estimate first remains about the same, because
it focuses on the
other features of the input. However,
as the simulated query enters the
malicious regime with a huge upper bound cost, the ML predicted cost
rapidly increases. That is, for malicious queries which
otherwise look normal but have huge actual cost and upper bound cost,
the ML prediction is also huge and will thus prevent the system
from executing such queries.
How the ML estimate
increases with the upper bound cost depends on the specific machine learning
model used. For example the plateaus at 
Figure~\ref{fig:bound} and Figure~\ref{fig:boundY} come from the choice of gradient boosted trees~(\textsf{GBR})
which ignore further increases in the upper bound cost.
The estimate reaches a final plateau because this algorithm is not good at extrapolating values outside of ranges of the training set.
The machine learning model can be tailored to the provider's risk preferences.
For instance, we could create a model where the output could be set to increase with  the static analysis upper bound.

The ML framework is flexible. A more conservative
provider has several options for further increasing the robustness of
the ML estimated cost. The first is to manually increase the contribution
of the upper bound cost. In the limit, the provider even has the option to
use only the upper bound cost
if no underestimation in cost can be tolerated.
Alternatively, the ML model can be trained with both typical
and malicious queries. This means 
\remove{demonstrates the relationship between the ML estimate and the static analysis estimate. 
Both GitHub and Yelp, the ML estimate maintains an uptrend with the static analysis estimate.
The Yelp estimates have some plateauing but we believe this can be attributed to the nature of the Yelp data, namely that the data has significant clustering and the static analysis radically overestimates queries with larger costs.
Nevertheless, because of the uptrend, we believe that the ML approach is making sensible estimates and have some robustness against pathological queries.}
inclusion of a wider collection of query examples in the data, including
instances of malicious queries. While several enhancements to robustness
against malicious queries are possible, our goal here was to
demonstrate that such robustness is already inherent to the ML
paradigm.

 \section{Threats to validity and discussion}
\label{sec:discussion}

\paragraph{Internal validity}

The first threat comes from the realism of the test datasets.
As far as we know, at the time of writing, there are no publicly available datasets of real-world production workloads for GraphQL 
queries and responses that we could use to test our framework directly.
The generated dataset available at~\cite{artifact} was small in size (10,000 query-response pairs) and not sufficient for ML purposes, 
instead we used the tooling provided to generate a much larger dataset (100,000 query-response pairs) for our experiments.
These queries were issued to GitHub and Yelp, which are two publicly available commercial APIs so they correspond with real-world data.
While we were able to produce good estimates for these datasets, it is important to keep in mind that randomly generated queries may not be representative of queries that human users may realistically design and run.
Moreover, the paper proposes a framework that should be trained by the service provider specifically for its system.
The service provider~(or gateway provider) can use automatically generated queries to bootstrap the system, but it can use also actual traffic for the training.

The second threat is due to the number of datasets used in the study.
We used two services, GitHub and Yelp, for the evaluation and we observe considerable differences in the data.
These differences have an impact on the quality of the estimate, but it still provides an improvement compared to the state-of-the-art.
This suggests that each API provider will need to tune the learners to address the specific characteristics of their system.

\paragraph{External validity}
An external threat to validity is the viability and persistence of the predictions over time. 
In general, the dynamics of actively managed data sources often change and evolve through time. 
They develop different morphology and structure and the data itself may also change.
This fact suggests that someone cannot expect persistently accurate predictions from a learning algorithm, meaning that the training of the learner needs to be updated with the new dynamics, if not continuously within a reasonable time frame.

Lastly, we need to consider what happens when our learners encounter malicious queries.
In Section~\ref{sec:robustness}, we tried to recreate this scenario by simulating pathological queries whose actual cost matches their upper bound cost.
We observe some robustness against malicious queries and this can be attributed to the fact that we use the static analysis as a feature in our algorithm.
Moreover, the more malicious queries the system receives, the better it becomes at recognizing them if they are used for retraining.
We also showed in Section~\ref{sec:practicality} that the underestimation of some query costs was compensated by the overestimation of some others.

 \section{Related Work}\label{sec:related}

Section~\ref{sec:intro} discusses previous work related to GraphQL cost estimation such as a query analysis that probes the service backend~\cite{hartig2018semantics} and an analysis that computes an upper bound~\cite{cha2020principled}.
This section discusses the broader context of ML techniques for cost estimation.

Our work is an instance of \textit{machine learning for code} (ML for
code).
ML for code has been extensively studied in the software engineering
community \cite{hindle_et_al_2012, rahman_palani_rigby_2019,allamanis_et_al_2018},
including for optimizing computational performance~\cite{wang_oboyle_2018}.
There are several works that use code as input, usually in the
form of a token sequence, and then train ML models for a variety of
tasks (for example, code completion). To the best of our knowledge, our
work is the first to apply ML to GraphQL.
Cummins et al.\ first train a language model
on OpenCL source code, then train a deep learning classifier to pick
performance optimizations~\cite{cummins_et_al_2017}. 
We focus on GraphQL instead of OpenCL and
predict performance instead of picking an optimization.
Our work obtains some of its features for code from a graph neural
network (GNN). Using GNNs for code was pioneered by Allamanis et
al.~\cite{allamanis_brockschmidt_khademi_2018}, who represented C\#
code as a multi-graph with edges for syntax, data flow, and control
flow. They used their model for variable name prediction tasks. In
contrast, we focus on GraphQL and predict query cost.

The database community has also studied \textit{query performance
  prediction} (QPP).
Akdere et al. use support vector machines and
other ML techniques to predict the performance of
relational database queries~\cite{akdere_et_al_2012}.  
Besides hand-crafted features such as
operator occurrence counts and query optimizer estimates, they propose
stacking models for individual relational operators to obtain a model
for a composite query.
Marcus and Papaemmanouil use similar
features and a similar stacking idea, but with deep learning, thus
enabling end-to-end learning with back-propagation into earlier
stacked layers~\cite{marcus_papaemmanouil_2019}. In contrast to both of these works, our work targets
GraphQL, which is more join-heavy, and does not require query
optimizer estimates.
Hasan and Gandon use support vector machines
and $k$-nearest neighbors for QPP for SPARQL~\cite{hasan_gandon_2014}. 
They use hand-crafted
features including operator occurrence counts, tree depth and size,
and graph edit distance to similar queries. While SPARQL has some
similarities to GraphQL, they differ substantially. Furthermore, our
work also uses a sound conservative upper bound on query cost as well
as graph neural network features.

 \section{Conclusion}
\label{sec:conclusion}

This paper proposes a methodology for using ML to estimate the cost of GraphQL queries.
We experimentally show that our ML approach outperform the leading existing static analysis approach using two commercial GraphQL APIs, namely GitHub and Yelp.
We believe that an ML approach to query complexity estimation can be useful for both API providers and clients.
API providers benefit by allowing them to loosen cost thresholds and better provision server resources, while clients benefit by allowing them to better understand the costs of their queries and what is allowable within their rate limits.
In addition, our approach can be used in conjunction with other types of analyses to create an overall more robust API management system.
 
\balance
\bibliographystyle{acm-reference-format}
\bibliography{graphql}

\end{document}